\documentclass[12pt,preprint]{aastex}
\newcommand{\eqref}[1]{(\ref{#1})}
\newcommand{\Log}{{\rm Log}}
\newcommand{\ba}{{\bf a}}
\newcommand{\bb}{{\bf b}}
\newcommand{\bc}{{\bf c}}
\newcommand{\beh}{\hat{\bf e}'}
\newcommand{\bq}{{\bf q}}
\newcommand{\qref}{q_{\rm ref}}

\newcommand{\br}{{\bf r}}
\newcommand{\brho}{{\bf \rho}}
\newcommand{\bs}{{\bf s}}
\newcommand{\bt}{{\bf t}}
\newcommand{\bu}{{\bf u}}
\renewcommand{\bv}{{\bf v}}
\newcommand{\bvh}{\hat{\bf v}_\perp}
\newcommand{\Wt}{\widetilde W}
\newcommand{\bx}{{\bf x}}
\newcommand{\bzero}{{\bf 0}}
\newcommand{\half}{\frac{1}{2}}
\newcommand{\rF}{r_{\rm F}}
\newcommand{\spar}{s_{\parallel}}
\newcommand{\sperp}{s_{\perp}}

\shorttitle{Flux scintillation}
\shortauthors{Goodman and Narayan}

\begin{document}

\title{Fitting Formula for Flux Scintillation of Compact Radio Sources}


\author{J. Goodman}
\affil{Princeton University, Princeton, NJ 08544}
\email{jeremy@astro.princeton.edu}

\and

\author{R. Narayan}
\affil{Harvard-Smithsonian Center for Astrophysics, Cambridge, MA 02138}
\email{narayan@cfa.harvard.edu}


\begin{abstract}
We present a fitting function to describe the statistics of flux
modulations caused by interstellar scintillation.  The function models
a very general quantity: the cross-correlation of the flux observed
from a compact radio source of finite angular size observed at two
frequencies and at two positions or times.  The formula will be useful
for fitting data from sources such as intra-day variables and
gamma-ray burst afterglows.  These sources are often observed at
relatively high frequencies (several gigahertz) where interstellar
scattering is neither very strong nor very weak, so that asymptotic
formulae are inapplicable.
\end{abstract}

%% subject header.  Objects should be in the appropriate "individual"
%% headers (e.g. quasars: individual, stars: individual, etc.) with the
%% additional provision that the total number of headers, including each
%% individual object, not exceed six.  The \objectname{} macro, and its
%% alias \object{}, is used to mark each object.  The macro takes the object
%% name as its primary argument.  This name will appear in the paper
%% and serve as the link's anchor in the electronic edition if the name
%% is recognized by the data centers.  The macro also takes an optional
%% argument in parentheses in cases where the data center identification
%% differs from what is to be printed in the paper.


% Read in the section ``Introduction''
%\include{intro}
\section{Introduction}

Ever since its initial discovery in the signals received from radio
pulsars, interstellar scattering and scintillation has proved to be a
useful tool in radio astronomy (see Rickett 1990, 2001; Narayan 1993;
Hewish 1993 for reviews).  It provides unique information on
small-scale turbulent fluctuations in the ionized interstellar medium (e.g.,
Armstrong, Cordes \& Rickett 1981; Wilkinson, Narayan \& Spencer 1994)
and on the sizes of compact radio sources.  The latter application has
contributed to progress in several areas of astrophysics: radio
pulsars (e.g., Roberts \& Ables 1982; Cordes, Weisberg \& Boriakoff
1983; Wolszczan \& Cordes 1987; Gwinn et al. 1997), intra-day
variables (e.g., Rickett et al. 1995, 2001; Lovell et al. 2003) and
gamma-ray burst afterglows (e.g., Goodman 1997; Frail et al. 1997;
Taylor et al. 1997).

The fundamental quantity in scintillation theory is the correlation of
flux variations of a distant source observed at two positions
separated by transverse distance $r$ on the observer plane.  Since
each flux is the square of the local electric field, the quantity of
interest is a fourth order correlation of the electric field (see
eq. \ref{Gamma4def}).  To interpret scintillation data, we need to
calculate the expectation value of this fourth moment as a function of
the strength of the turbulent fluctuations in the scattering medium,
the shape and size of the radio source, the observing frequency, etc.

The scattering of radio waves in the interstellar medium may be
characterized by two dimensionless parameters (eq. \ref{pwrsf}): $U$
which measures the strength of the scattering, and $\alpha$ which
describes the power spectrum of the fluctuations.  The turbulence in
the interstellar medium\footnote{We use the term ``turbulence'' as a
short-hand for both true dynamic turbulence as well as passive density
irregularities that may not exhibit turbulent dynamics} appears to be
well-described by a Kolmogorov scaling, which corresponds to
$\alpha=5/3$.  Hence, the scattering medium along any line of sight is
determined by just $U$; however, $U$ varies from one direction to
another, and also with frequency in a given direction (see
eq. \ref{Ulambda}).  Analytical results are available for the flux
correlation function in the limit of both very weak scattering ($U\ll
1$) and very strong scattering ($U\gg1$) (e.g., Goodman \& Narayan,
1995; Rickett 1990; and references therein).  In the latter regime,
the scintillation is known to occur on two very different scales,
determined by diffractive and refractive effects.

While a great deal of interesting work on interstellar scintillation
has been done using asymptotic results, many observations correspond
to the difficult intermediate regime where $U$ is neither $\ll1$ nor
$\gg1$; no analytical results are available in this transition regime.
For typical high lines of sight through the interstellar medium at
high Galactic latitudes, the transition regime occurs at radio
frequencies $\sim5-10$ GHz, a frequency of much interest for both
intra-day variables and gamma-ray burst afterglows (but less so for
pulsars which, because of their steep radio spectra, are usually
observed at lower frequencies where asymptotic strong scattering
results apply).  Some very approximate formulae have been proposed for
the transition regime (e.g., Walker 1998).  However, for accurate
results, one has to resort to numerical computations of the
scintillation correlation function, which is technically challenging
and has been rarely attempted.

We describe in this paper numerical computations of the scintillation
correlation function for a wide range of values of the scattering
parameter $U$, spatial separation $r$, source size $r_s$ (defined in
eqs. \ref{gaussian}--\ref{rsdef}) and frequency difference $\eta$
(defined in eq. \ref{etadef}).  Using the numerical results we have
developed a fitting function for the flux correlation that is valid
for all values of $U$.  The function asymptotes to the appropriate
analytical results in the limits $U\ll1$ (very weak scattering) and
$U\gg1$ (very strong scattering) and agrees well with numerical
results in the transition regime in between.

In \S\ref{sec:screen} we introduce the basic fourth order moment of
interest to flux scintillation observations, and in \S
\ref{sec:numerical} we explain how we numerically compute this
quantity.  In \S\ref{sec:fitting} we present our fitting formula,
which agrees well with the numerical results.  We conclude in
\S\ref{sec:summary} with a brief summary.  We present in
Appendices \ref{AppA}-\ref{AppC} technical details, including some
asymptotic results useful as checks of our fitting functions, and a
discussion of parabolic arcs in secondary spectra.

% Read in ``Thin-screen theory''
%\include{screen}
\section{Thin-screen theory}\label{sec:screen}

In the thin-screen approximation, one imagines that the turbulence
affecting radio-wave propagation is concentrated in a narrow layer
perpendicular to the line of sight.  This is sometimes not far from
the truth, although it would usually be more accurate to assume a
number of scattering screens; but the opposite limit of homogeneously
distributed turbulence is probably the exception rather than the rule.
The observer lies in a parallel plane at distance $z_{\rm screen}$
from the screen, and the source is behind the screen at distance
$z_{\rm source}$ from it.  Waves propagate freely from the source to
the screen, where they suffer a phase shift $\phi(\bx)$ that depends
upon the two-dimensional position $\bx$ within the screen, and thence
travel freely again to the observer's plane.  The distortion of the
phase fronts at the screen leads to modulations of the flux on the
observer's plane, through a combination of refractive focusing and
diffractive interference.  It is useful to define an \emph{effective
distance}
\begin{equation}\label{zdef}
z\equiv\rm\left(\frac{1}{z_{\rm screen}}+\frac{1}{z_{\rm source}}\right)^{-1}
\end{equation}
from observer to screen.  This allows us to analyze the problem as if
the source were infinitely distant and the wavefronts were planar before
encountering the screen.

We follow the notation of Goodman \& Narayan (1989, henceforth GN89),
who discuss the mutual coherence function
\begin{equation}\label{Gamma4def}
\Gamma_4(\bb;\br;\nu_1,\nu_2)\equiv\left\langle
E\left(\half\bb,\nu_1\right)E^*\left(-\half\bb,\nu_1\right)
E\left(\br+\half\bb,\nu_2\right)E^*\left(\br-\half\bb,\nu_2\right)\right\rangle\,,
\end{equation}
where $E(\br,\nu)$ is the electric field due to a source of unit
strength measured at vector position $\br$ on the observer's plane (a
plane perpendicular to the line of sight to the source) at radio
frequency $\nu$.  As written above, $\Gamma_4$ is useful for studying
the correlation between measurements of visibility made with two
two-element interferometers having the same baseline $\bb$ but phase
centers separated by $\br$, and operating at different frequencies
$\nu_1$ and $\nu_2$.  In this paper, we are interested in flux
correlations rather than visibilities, so we set $\bb\to 0$.
Positions $\bzero$ and $\br$ then represent two positions at which the
flux is measured.  The measurements may be done by two telescopes
simultaneously, or by the same telescope at different times.  In the
latter case, the interstellar turbulence responsible for phase changes
in $E$ is assumed ``frozen'' on timescales of interest, so that
$\br=\bv_\perp t$, where $\bv_\perp$ is the effective transverse
velocity of the line of sight through the interstellar medium.  The
angle brackets denote an ensemble average over realizations of the
turbulence.

Appendix A derives formula \eqref{G4final} for the flux correlation
$\Gamma_4(\bzero;\br;\nu_1,\nu_2)$ in terms of the phase structure
function
% The following formula...
%\[
%D(\Delta\bx)\equiv\left\langle\phi(\bzero,\lambda)\phi(\Delta\bx,\lambda)
%\right\rangle
%\]
% ...should be replaced by
\[
D(\Delta\bx)\equiv\left\langle
\left[ \phi(\Delta\bx,\lambda) -\phi(\bzero,\lambda) \right]^2
\right\rangle
\]
evaluated at the geometric mean wavelength
\begin{equation}\label{lambda}
\lambda \equiv \sqrt{\lambda_1\lambda_2}
\end{equation}
of the two observing wavelengths $\lambda_{1,2}\equiv c/\nu_{1,2}$,
and in terms of a Fresnel scale defined with the
\emph{arithmetic} mean,
\begin{equation}\label{rFdef}
\rF\equiv\left[\frac{(\lambda_1+\lambda_2)z}{4\pi}\right]^{1/2}\,,
\end{equation}
and a dimensionless frequency difference
\begin{equation}\label{etadef}
\eta\equiv\frac{\lambda_2-\lambda_1}{\lambda_1+\lambda_2}~:~~ 0\le|\eta|<1\,.
\end{equation}

In this paper, we restrict our attention to power-law structure functions
\begin{equation}\label{pwrsf}
D(\Delta\bx)= U\left(\frac{\Delta\bx}{\rF}\right)^{\alpha}\,,
\end{equation}
and unless otherwise specified, we assume $\alpha=5/3$ (the
``Kolmogorov'' value).  Thus $U$ is a dimensionless number describing
the strength of scattering; it depends upon the observing wavelength
(see eq. \ref{Ulambda}) as well as the intrinsic amplitude of the
turbulent electron-density power spectrum, since $D\propto\lambda^2$
as a consequence of the plasma dispersion relation.  Henceforth we
choose units of length such that $\rF\equiv1$.  Also, we normalize the
mean flux to unity at all frequencies, $\langle
F(\br,\nu)\rangle=\langle|E(\br,\nu)|^2\rangle=1$.  Writing $W$ rather
than $\Gamma_4$ for the flux correlation, we have
\begin{eqnarray}\label{Wdef}
W(\br,\eta)&\equiv&
\Gamma_4({\bf 0}; {\bf r}; \nu_1, \nu_2) \equiv
\left\langle
\left|E(\bzero,\nu_1)\right|^2\left|E(\br,\nu_1)\right|^2\right\rangle
=\int\frac{d\bq}{(2\pi)^2}\,\exp(i\br\cdot\bq)
\Wt(\bq,\eta),\\[1ex]
\label{Wtdef}
\Wt(\bq,\eta) &=& \exp\left[-\half F(\bq,\eta)\right]
\int d\bs\,\exp(i\bq\cdot\bs)
\exp\left[-\half G(\bs,\bq,\eta)\right],\\[2ex]
\label{Fdef}
F(\bq,\eta)&\equiv& U\left[\frac{(1-\eta)^{\alpha+2}+
(1+\eta)^{\alpha+2}}{1-\eta^2}\right]\left|\bq\right|^\alpha
,\\[2ex]
\label{Gdef}
G(\bs,\bq,\eta)&\equiv& U\left[
\left|\bs-\eta\bq\right|^\alpha+\left|\bs+\eta\bq\right|^\alpha-
\left|\bs-\bq\right|^\alpha-\left|\bs+\bq\right|^\alpha\right]\,.
\end{eqnarray}
The quantity $\Wt(\bq,\eta)$ will be referred to as the ``cross spectrum.''
For $\eta=0$ it reduces to the two-dimensional spatial power spectrum of the
flux, $\Wt(\bq)$.

We know no closed-form results for the integral \eqref{Wtdef} except
at uninteresting values of $\alpha$, so it is necessary to evaluate
the integral numerically in general.  However, approximate analytic
expressions can be obtained in the limits $U\ll 1$ (weak scattering)
and $U\gg 1$ (strong scattering), and these are given in Appendix B.
They are useful to check the accuracy of the numerical evaluation of
the integral \eqref{Wdef}, and also to guide the choice of functional
form for the fitting functions presented in \S4.

% Read in ``Numerical methods''
%\include{numer}
\section{Numerical methods}\label{sec:numerical}

The integral \eqref{Wtdef} for the flux cross spectrum
$\Wt(\bq,\eta)$ proved to be
challenging to estimate accurately over the full ranges of $U$, $q$, and
$\eta$ required.  It is two-dimensional 
and extends to infinity.  Worse, it is only conditionally
convergent since $G(\bs,\bq,\eta)\to 0$ as $s/q\to\infty$.
And the regions of the $\bs$ plane that dominate the integral vary with
the scattering regime.

We experimented with several approaches before settling on the
following.  The independent variable $\bs$ is resolved into its
components parallel and perpendicular to $\bq$: $\spar$ \& $\sperp$.
The inner integration is taken over $\spar$ and is calculated by a
version of steepest descent in the complex plane $\spar\equiv z\equiv
x+iy$.  That is, at each value of $\sperp$ required by the quadrature
scheme for the outer integral, our code estimates
\begin{equation}\label{contour_int}
\int\limits_{C} \exp\left[iqz~+~G(z,\sperp,q,\eta)\right]\,dz,
\end{equation}
where $C$ is a contour which in the first instance is the real axis
[$y\equiv0$, $x\in(-\infty,\infty)$] but which is moved into the upper
half plane, where the first exponential in \eqref{contour_int}
vanishes as $y\to\infty$ since $q\equiv|\bq|>0$.  The contour must
not, however, be dragged across any of the singularities of
\begin{eqnarray}\label{Gsing}
G(z,\sperp,q,\eta)&=& U\left\{\left[(z-\eta q)^2+\sperp^2\right]^{\alpha/2}
+\left[(z+\eta q)^2+\sperp^2\right]^{\alpha/2}\right.
\nonumber\\
&\phantom{=}&
\left.-\left[(z- q)^2+\sperp^2\right]^{\alpha/2}
+\left[(z+ q)^2+\sperp^2\right]^{\alpha/2}\right\},
\end{eqnarray}
which has algebraic branch points at $y=\sperp$ and $x\in\{\pm
q,\pm\eta q\}$.  A branch cut must be drawn from each point.  Unless
$4\alpha$ is integral, these cuts must extend to infinity rather than
join the branch points to one another.  We chose to draw the cuts
upward parallel to the imaginary axis.  This is convenient because
$[(z-a)^2+\sperp^2]^{\alpha/2}$ is evaluated as
\[
\exp\left\{(\alpha/2)\Log
\left[(z-a)^2+\sperp^2\right]\right\}\,,
\]
and complex library functions of scientific programming languages usually
put the branch cut of $\Log$ on the negative real axis, which corresponds
to the cuts in the $z$ plane that we have described.  For pedagogical reasons,
we wrote the code in C, although that language has only recently 
supported complex arithmetic.  Recent
versions of the GNU {\tt gcc} compiler and mathematical library functions
performed very reliably and were almost as convenient to use as their
\textsc{Fortran} counterparts.

% This figure shows a typical integration contour.  It may be dispensible.
%\clearpage
\begin{figure}[p]
\epsscale{0.80}
%\plotone{contourfig.eps}
\plotone{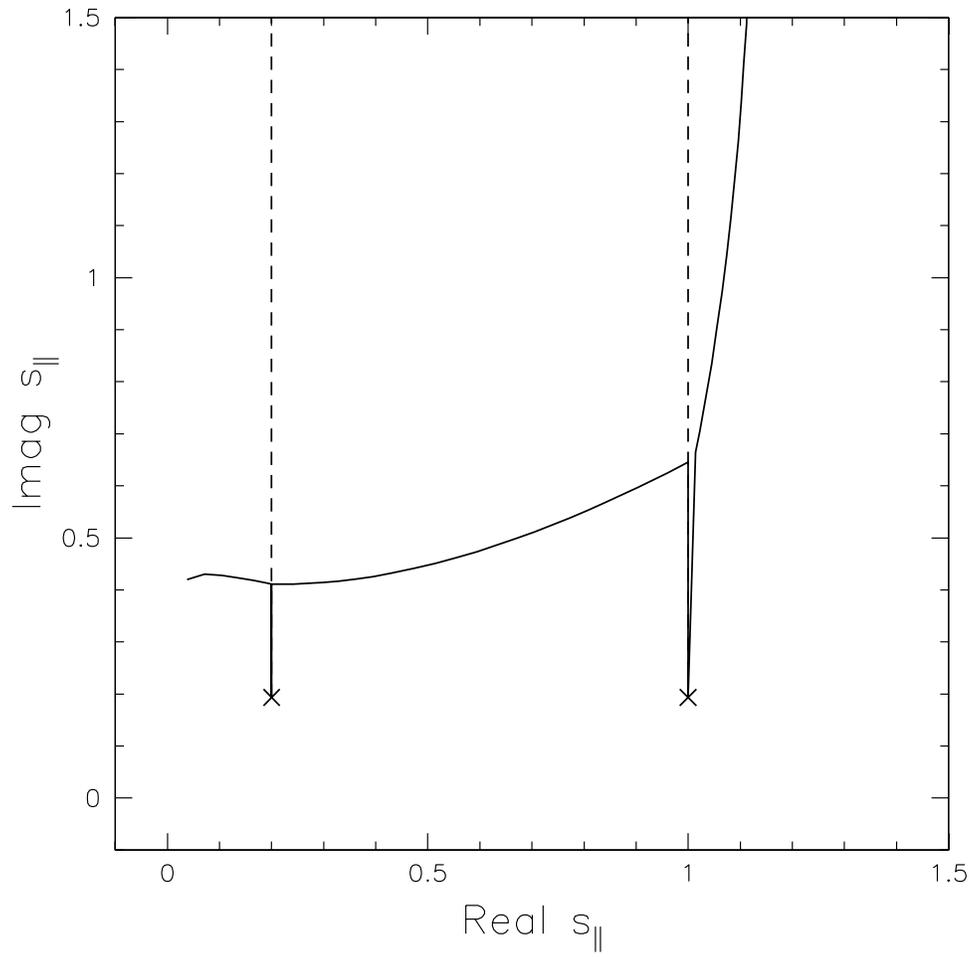}
\caption{Integration contour (solid line) and branch cuts (dashed)
for $\alpha=5/3$, $U=4$, $\eta=0.2$, $q=1$, $\sperp=0.1933$.
\label{contourfig}}
\end{figure}
%\clearpage
A typical contour of integration is shown in Figure~\ref{contourfig}.
Integration starts from $x=0$, since the complete integral
\eqref{contour_int} is twice its righthand half, but at $y>0$, since there
is no singularity to pin the contour to the origin.
The contour is drawn adaptively, differentiating the argument of the
exponential at each step to determine the direction in which its real part
becomes negative most rapidly and its imaginary part is constant,
until a branch cut is encountered.  The contour then follows
the lefthand side of the cut downward to the branch point,
circumnavigates it, and then continues by steepest descent.
Special care must be taken when $q$
is small since there can be near-cancellation between the contributions from
the cuts at $x=\eta q$ and $x=q$; this is handled by applying a
Romberg quadrature scheme to the sum of the integrands
at $z=(\eta q-0^+,y)$ and $(q-0^+,y)$.  Care must also be taken
when $|\bs|\gg q$ because the terms of $G$ nearly cancel; this is handled
by expanding $G$ in powers of $q/z$.

The numerical estimate of \eqref{contour_int} becomes the integrand
for integration over $\sperp$.  This integrand 
decreases exponentially at large $\sperp$ because
the branch points of the inner integral move far from the real axis
where their contribution is suppressed by $\sim\exp(-q\sperp)$.  We
use a straightforward midpoint quadrature along the real axis in the
auxiliary variable $t\equiv\sinh^{-1}(\sperp/s_0)$, where $s_0$ is a
scale chosen according to the values of $q$ and $U$.  As is well
known, numerical quadratures of smooth functions over infinite or periodic
domains with schemes that use uniform steps and weights,
such as midpoint or Simpson's rule, converge faster than any power of
stepsize.  So there is no point in using a higher-order method for the
outer integral.

The present method for evaluating $\widetilde W(q)$
is efficient for moderate to large values of $U$
and $q$ because the $\spar$ integrand decreases rapidly along the
steepest-descent path, but it has difficulty with very small $Uq^\alpha$.
Fortunately, the analytic weak-scattering approximation is then very
accurate, so the code uses that approximation when
$Uq^\alpha<10^{-4}$.  One test of the code is that the numerical
results match smoothly onto the analytic ones in the weak-scattering
regime.  We have tested the code also against the asymptotic results
of the previous section for strong scattering, $U\gg1$.  And we have
tested it against an old code valid only for $\eta=0$ that uses a
completely different strategy (based on integrating $\bs$ in polar
rather than cartesian coordinates; see Goodman \& Narayan 1985).

After tabulating $\widetilde W(\bq;U,\eta)$ for given values of $U$
and $\eta$, we obtain the flux correlation function $W(\br)$ by
numerical calculation of the Hankel transform
\begin{equation}\label{Hankel}
W(U,r,\eta)= \frac{1}{2\pi}\int\limits_0^\infty J_0(qr)\widetilde
W(q;U,\eta)
\,q dq\,,
\end{equation}
where $W(\br)$ and $\widetilde W(\bq)$ depend on the absolute values
of $\br$ and $\bq$ only.  (This would not be true of the spectrum of
phase fluctuations on the scattering screen were anisotropic.)
Equation (\ref{Hankel}) is for a point source.  In the general case,
when the source observed at position $\bf 0$ has a circularly
symmetric normalized intensity profile $S_1(r)$ with root mean square
size $r_{s1}$ and that observed at position $\bf r$ has profile
$S_2(r)$ with size $r_{s2}$, the flux correlation function becomes
\begin{equation}
W(U,r,r_{s1}, r_{s2},\eta)= \frac{1}{2\pi}\int\limits_0^\infty J_0(qr)
\widetilde S_1(q) \widetilde S_2(q) \widetilde W(q;U,\eta)
\,q dq\,,
\label{finitesource}
\end{equation}
where $\widetilde S_1(q)$ and $\widetilde S_2(q)$ are the Hankel
transforms of the source profiles:
\[
\widetilde S_{1,2}(q) = \int\limits_0^\infty J_0(qr) S_{1,2}(r) rdr.
\]
In this paper, we consider two different source profiles, gaussian and
tophat.  The normalized profiles are
\begin{eqnarray}
\mbox{Gaussian}&:& S(r)= \frac{1}{\pi r_s^2}e^{-(r/r_s)^2}\,;
\quad\widetilde S(q)= e^{-(q r_s/2)^2}\label{gaussian} \\
\mbox{Tophat}  &:& S(r)= \frac{1}{\pi r_s^2}H(r_s-r)\,;
\quad\widetilde S(q)= \frac{2}{q r_s} J_1(q r_s)\,, \label{tophat}
\end{eqnarray}
where $r_s$ is a measure of the source size, and $H(x)=1$ if $x>0$,
$H(x)=0$ if $x<0$ (Heaviside function).  Note that the root mean
square radius of the source $r_{\rm rms}$ is related to the nominal
size $r_s$ as follows:
\begin{equation}\label{rrms}
r_{\rm rms} = r_s ~~({\rm Gaussian}), \qquad
r_{\rm rms} = r_s/\sqrt{2} ~~({\rm Tophat}).
\end{equation}
The ratio $r_{\rm rms}/r_s$ has an influence on the constants $C_1$
and $C_2$ defined in \S\ref{weakscatt}.

% Read in ``Fitting formulae''
%\include{fitting}
\section{Fitting Function}\label{sec:fitting}

In developing a fitting function for the correlation $W(U,r,r_{s1},
r_{s2},\eta)$ in equation (\ref{finitesource}), we have found that it
is not necessary to consider the dependence of $W$ on the two sources
sizes $r_{s1}$ and $r_{s2}$ independently.  Rather it is sufficient to
consider a single {\it effective size}
\begin{equation}\label{rsdef}
r_s \equiv \left(r_{s1}^2 + r_{s2}^2\right)^{1/2}.
\end{equation}
Thus, the most general scintillation flux correlation function that we
consider in this paper is $W(U,r,r_{s},\eta)$, where $U$ measures the
strength of the phase fluctuations at the scattering screen at the
geometric mean wavelength $\lambda$ (eqs. \ref{pwrsf} and
\ref{lambda}), $r$ is the distance in Fresnel units $r_F$
(eq. \ref{rFdef}) between the two points at which the fluxes are
measured, $r_{s}$ is the effective source size in Fresnel units as
projected at effective distance $z$ (eq. \ref{zdef}), and $\eta$ is
the dimensionless frequency difference between the two observations
(eq. \ref{etadef}).  All the results presented here are for a
Kolmogorov spectrum of fluctuations in the scattering screen
($\alpha=5/3$).  For this choice of $\alpha$, the parameter $U$ varies
with wavelength $\lambda$ as
\begin{equation}\label{Ulambda}
U(\lambda) = \left({\lambda\over\lambda_0}\right)^{(4+\alpha)/2},
\end{equation}
where $\lambda_0$ is the wavelength at which $U=1$.  Thus, the
scattering power of the screen is uniquely defined by the wavelength
$\lambda_0$.

Using the code described in the previous section, we have calculated
numerical values of $W(U,r,r_{s},\eta)$ over a wide range of values of
$U$, $r$, $r_s$ and $\eta$, and we have developed a fitting function
that approximates the numerical results.

\subsection{Mean Square Flux Variations for a Point Source}

\begin{figure}
%\epsscale{0.49} 
%\plotone{Ueta0r0rs0.ps}
\plotone{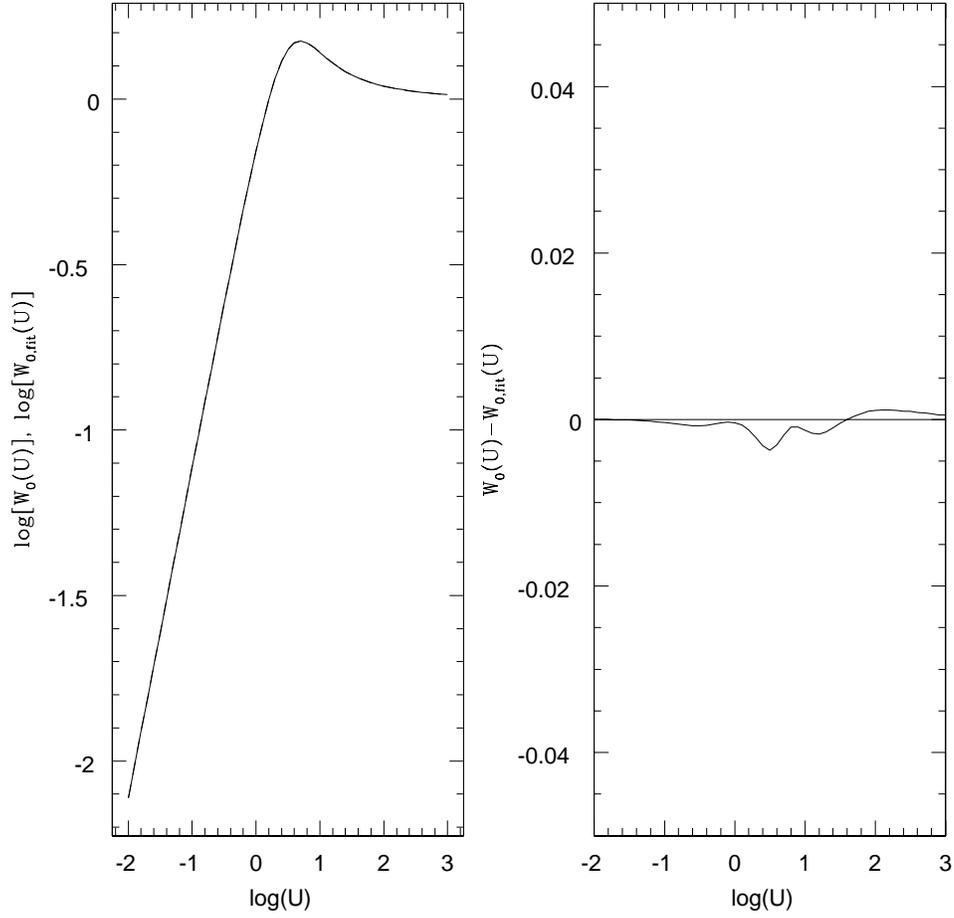}
\caption{Left: The solid line is the correlation $W_0(U)$ as
calculated numerically, and the dashed line is the fitting function
(hardly seen).  Right: The difference between the calculated
correlation and the fitting function.}
\label{figW0}
\end{figure}

In the limit when $r=r_{s}=\eta=0$, the quantity
\begin{equation}\label{W0def}
W_0(U) \equiv W(U, r=0, r_s=0, \eta=0)
\end{equation}
measures the mean square flux variations due to scintillation as a
function of the scattering strength $U$.  The panel on the left in
Figure \ref{figW0} shows a numerical evaluation of this function.
An accurate fitting function is given by
\begin{equation}\label{W0fit}
W_0(U) = {0.7729U\over 1+0.286U+0.0860U^2+0.0550U^3} +
{1+0.4760U^{-0.4}\over 1-1.64U^{-1}+10.1U^{-2}}\,,
\end{equation}
which has an error much less than 1\% over all values of $U$
(Fig. \ref{figW0}, right).
%\clearpage
%\clearpage
The function $W_0(U)$ crosses unity at
\begin{equation}
U_0 = 1.5874.
\end{equation}
We use this value of $U$ to distinguish between the two major regimes
of scattering.  When $U\leq U_0$, we consider that we are in the weak
scattering regime and seek to explain the scintillation properties
using a single scale, the weak scale.  For $U > U_0$, on the other
hand, we consider that we are in the strong scattering regime and
allow for two scales, one for diffractive scintillation and one for
refractive scintillation.  These are explained in the subsections
below.

\subsection{Weak Scattering ($U\leq U_0$)}\label{weakscatt}

In this regime, we write the fitting function for the general
correlation $W(U,r,r_s,\eta)$ in the form
\begin{equation}\label{weakfit}
W(U,r,r_{s},\eta) = W_0(U) F_r F_s F_\eta,
\end{equation}
where
\begin{eqnarray}
\label{Frdef}
F_r &=& \exp \left[-(r/R_1)^{5/3}\right],
\\ [1ex]
\label{Fsdef}
F_s &=& \left[1+(r_s/R_2)^{1.81}\right]^{-1}, \\ [1ex]
\label{Fetadef}
F_\eta &=& \left[1+a_1(U)\zeta^{5/6}+a_2(U)\zeta^2\right]^{-1},
\\ [1ex]
\label{zetadef}
\zeta &=& \eta/(1-\eta), \\ [1ex]
R_1 &=& R_3+3.54\eta^{1.12}/R_3^{1.09}, \\ [1ex]
R_2 &=& R_4+3.54\eta^{1.12}/R_4^{1.09}, \\ [1ex]
R_3 &=& \left( R_5^2+C_1 r_s^2\right)^{1/2}, \\ [1ex]
R_4 &=& C_2 R_5, \\ [1ex]
\label{R5weak}
R_5 &=& 1.15-0.260(U/U_0)^{0.730}, \qquad {\rm (weak~scattering~only)},
\\ [1ex]
a_1(U) &=& \left[1.01^2+(0.747U)^2\right]^{1/2}, \\ [1ex]
a_2(U) &=& \left[ \left(11.9U^{1.02}\right)^2+\left(5.37U^{2.4}\right)^2
\right]^{1/2}.
\end{eqnarray}

The three factors $F_r$, $F_s$ and $F_\eta$ describe the variation of
the flux correlation as a function of separation $r$, source size
$r_s$ and wavelength difference $\eta$.  Each factor is defined such
that it goes to unity in the limit when $r=r_s=\eta=0$, so that
equation (\ref{weakfit}) reduces to equations (\ref{W0def}) and
(\ref{W0fit}) in this limit.

There are two constants, $C_1$ and $C_2$, in the above formulae.  The
values of these constants depend on the shape of the source.  For the
gaussian and tophat source models defined in \S~3, we find
\begin{eqnarray}
{\rm Gaussian}&:& \quad C_1 = 1.37, \quad C_2 = 0.965, \\
{\rm Tophat}  &:& \quad C_1 = 0.829, \quad C_2 = 1.32.
\end{eqnarray}
All the other numerical constants in the fitting function are
independent of source shape.  We suspect that the above two source
models are sufficient for most applications --- the gaussian will
serve for the majority of sources and the tophat is suitable for
gamma-ray burst afterflows (Sari 1998).  In case one needs to consider
other source shapes, we note that $C_1$ and $C_2$ scale approximately
as
\begin{equation}\label{C1C2}
C_1 \approx 1.4 \left(\frac{r_{\rm rms}}{r_s} \right)^2, \qquad
C_2 \approx 0.95 \left(\frac{r_s}{r_{\rm rms}} \right),
\end{equation}
where $r_{\rm rms}$ is the root mean square size of the image.
Equation (\ref{rrms}) gives $r_{\rm rms}/r_s$ for the gaussian and
tophat models.  The scaling (\ref{C1C2}) is rather approximate because
the best-fit values of $C_1$ and $C_2$ depend on more than just the
second moment of the source.

\subsection{Strong Scattering ($U > U_0$)}\label{strongscatt}

When $U > U_0$ we are in the regime of strong scattering and the flux
variations have contributions from both diffractive and refractive
scintillation.  In the limit when $r=r_s=\eta=0$, we assume that the
mean square flux variations $W_0(U)$ consist of contributions $W_d(U)$
and $W_r(U)$, respectively, from diffractive and refractive
scintillation.  We take these contributions to be given by
\begin{equation}\label{Wdrdef}
W_d(U) = \left[{W_0(U)+1 \over 2}\right] , \qquad
W_r(U) = \left[{W_0(U)-1 \over 2}\right].
\end{equation}
The two quantities add up to give $W_0(U)$, as they should.  They are
also so defined that when $U\to U_0$, the refractive term vanishes,
i.e., at the boundary between weak and strong scattering, we have only
diffractive scintillation.  Thus, the fitting function will be
continuous across $U_0$ if we match the weak scattering function
(\ref{weakfit}) of the previous subsection with the diffractive term
in the function (\ref{Wstrong}) below.

With this motivation, we write the fitting function in the strong
scattering regime as
\begin{equation}
W(U,r,r_s,\eta) = W_d(U) F_r F_s F_\eta + 
W_r(U) F_{r,r} F_{s,r} F_{\eta,r},
\label{Wstrong}
\end{equation}
where the first term on the right is the diffractive term and the
second is the refractive term.  We take the factors $F_r$, $F_s$ and
$F_\eta$ in the diffractive term to have the same forms as in
\S~\ref{weakscatt}, with the sole exception that the scale $R_5$ is
now given by
\begin{equation}
R_5 = 0.964U^{-0.6}+\left(0.890-0.964U^{-0.6}\right)
(U_0/U)^{1.19}, \qquad {\rm (strong~scattering~only)}.
\label{R5strong}
\end{equation}
The two expressions (\ref{R5weak}) and (\ref{R5strong}) for $R_5$ are
equal at $U=U_0$.  Therefore, we have a perfect match at $U=U_0$
between the fitting functions for weak scattering and strong
diffractive scattering.

The second term on the right in equation (\ref{Wstrong}) describes
refractive scintillation.  Here we set
\begin{eqnarray}
\label{Frrdef}
F_{r,r} &=& \exp \left[-(r/R_6)^{7/3}\right], \\ [1ex]
\label{Fsrdef}
F_{s,r} &=& \left[1+(r_s/R_7)^{7/3}\right]^{-1}, \\ [1ex]
\label{Fetardef}
F_{\eta,r} &=& \left[1+a_3(U)\zeta^{5/6}+a_4(U)\zeta^2\right]^{-1}, \\ [1ex]
R_6 &=& R_8+15.1\eta^{1.48} , \\ [1ex]
R_7 &=& R_9+15.1\eta^{1.48} , \\ [1ex]
R_8 &=& \left( R_{10}^2+C_1r_s^2\right)^{1/2}, \\ [1ex]
R_9 &=& C_2 R_{10}, \\ [1ex]
R_{10} &=& 1.20-2.00U^{0.3}+1.73U^{0.6}, \\ [1ex]
a_3(U) &=& a_1(U_0)(U_0/U)^{0.567} , \\ [1ex]
a_4(U) &=& 8.38+\left[a_2(U_0)-8.38\right](U_0/U)^{0.933}.
\end{eqnarray}
The constants $C_1$ and $C_2$ have the same values as before.

\subsection{Comparison with Numerical Results}

\begin{figure}
\epsscale{0.49} 
%\plotone{Ueta0rrs0.w.ps}
%\plotone{Ueta0rrs0.1.ps}
%\plotone{Ueta0rrs0.2.ps}
%\plotone{Ueta0rrs0.s.ps}
\plotone{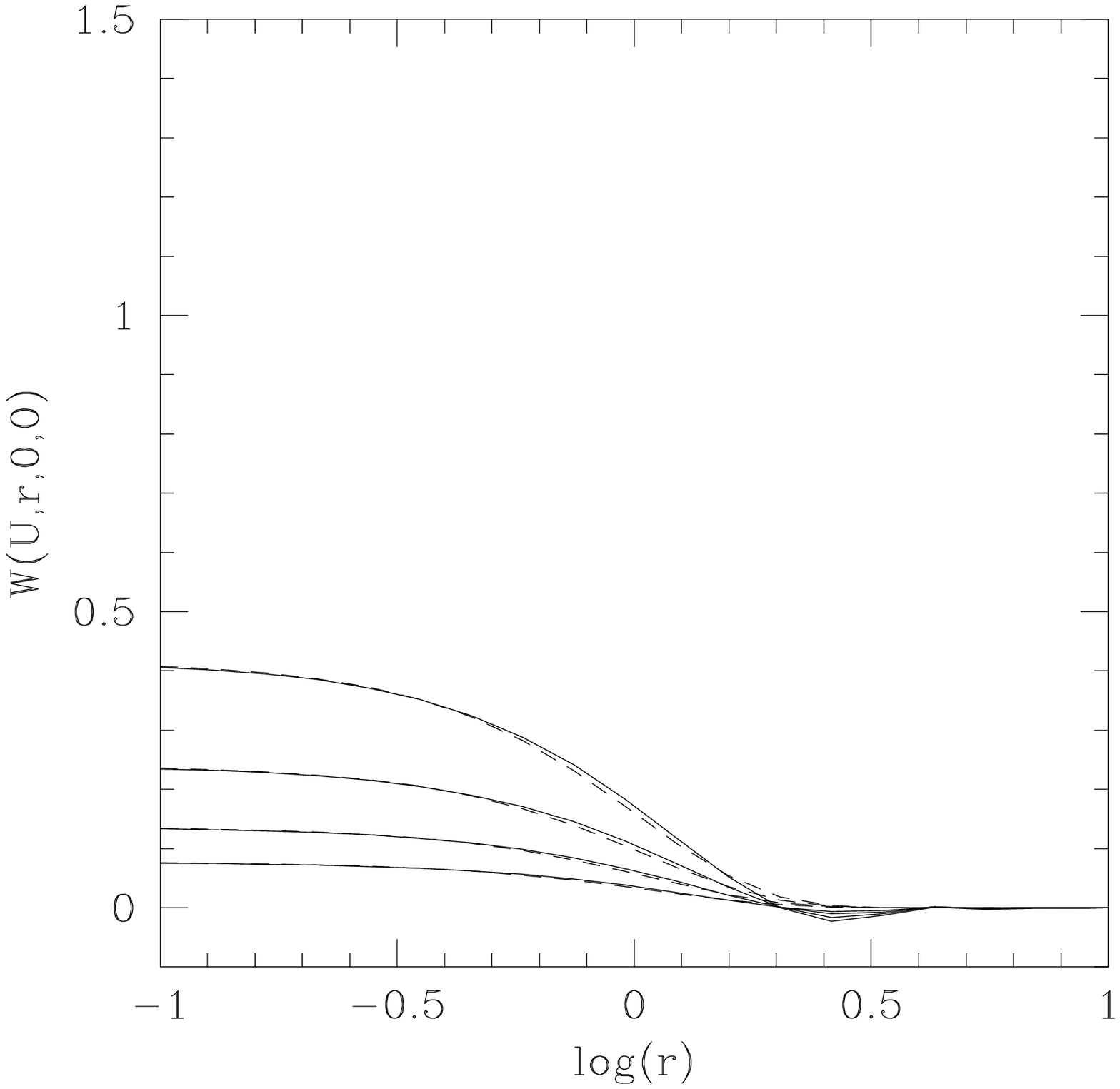}
\plotone{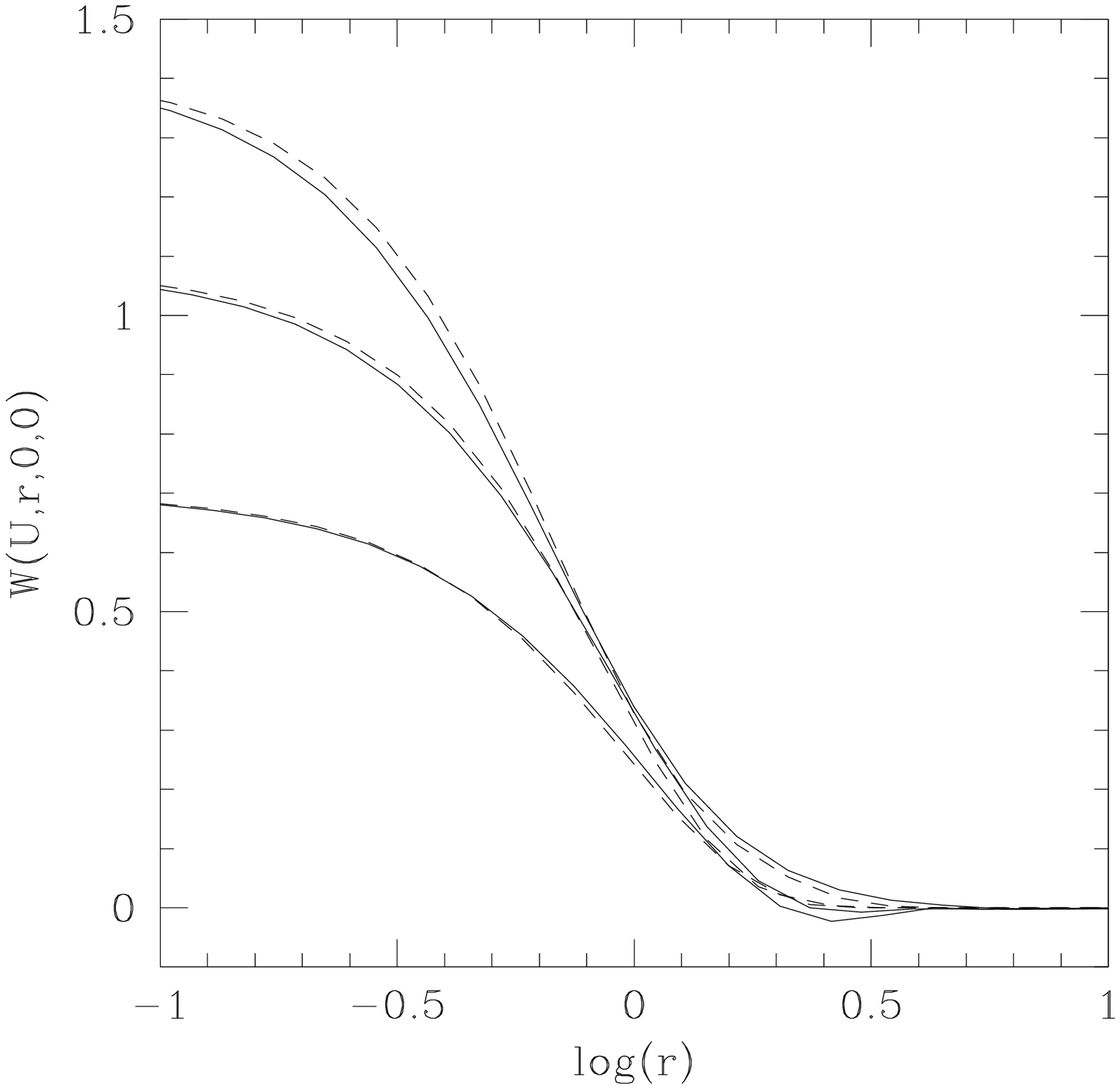}
\plotone{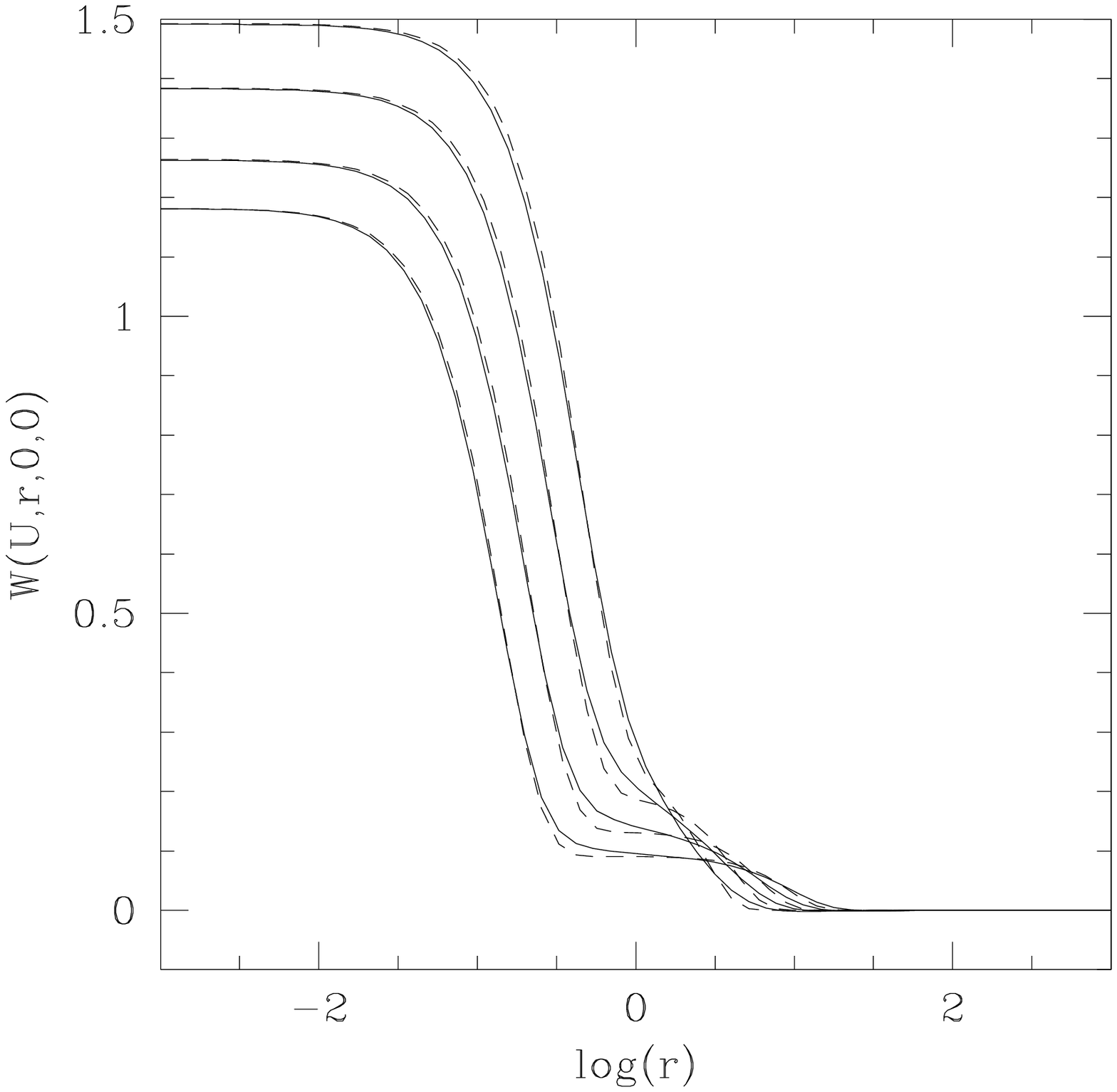}
\plotone{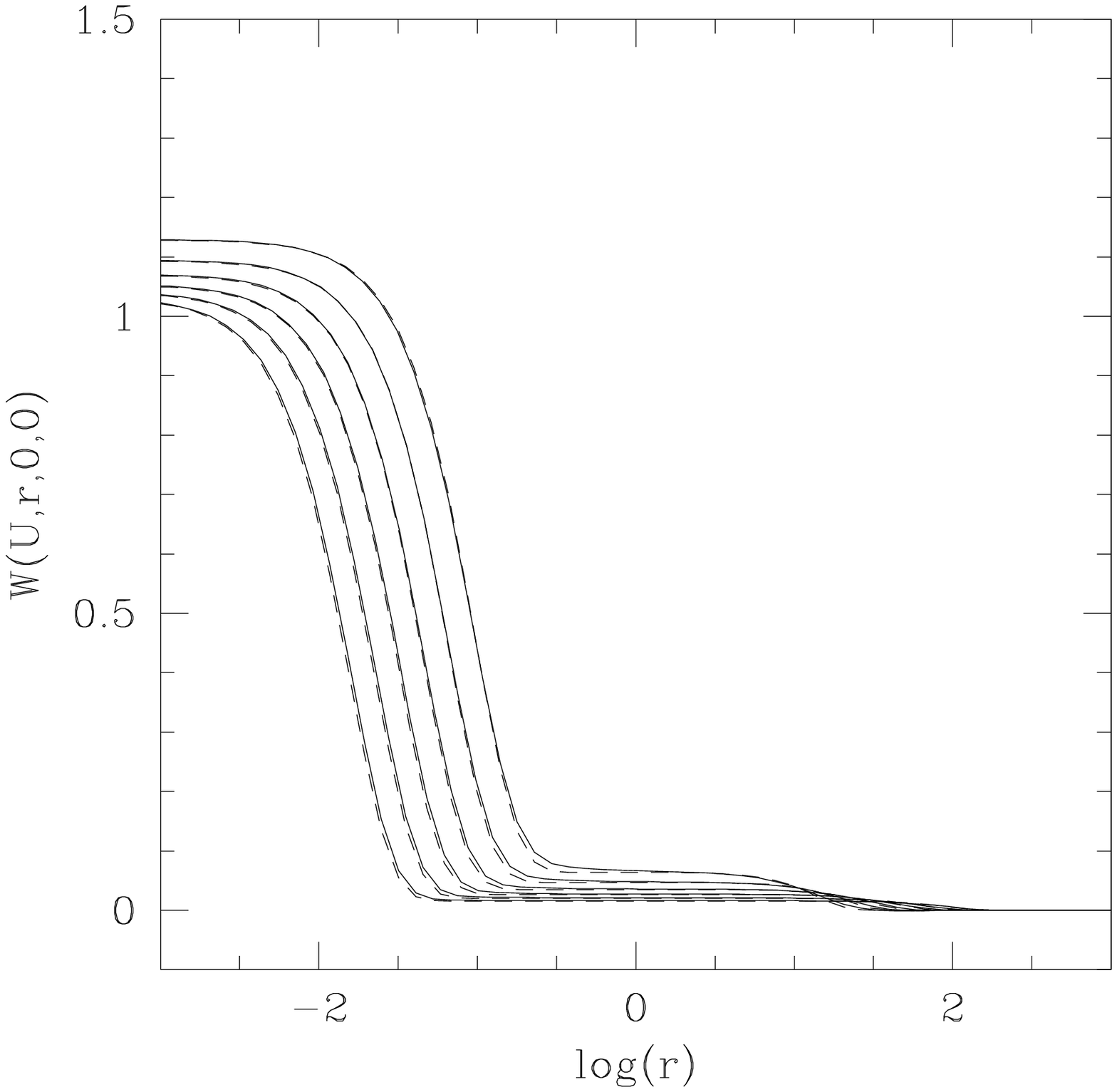}
\caption{Solid lines show the correlation as a function of separation 
$r$ for zero source size ($r_s=0$) and zero wavelength difference
($\eta=0$) as calculated numerically.  Dashed lines show the
corresponding fitting function.  Top Left: $\log(U) = -0.25, ~-0.5,
~-0.75, ~-1.0$ (from above).  Top Right: $\log(U) = 0.5, ~0.25, ~0$
(from above).  Bottom Left: $\log(U) = 0.75, ~1.0, ~1.25, ~1.5$ (from
above).  Bottom Right: $\log(U) = 1.75, ~2.0, ~2.25, ~2.5, ~2.75,
~3.0$ (from above).}
\label{figWr}
\end{figure}

\begin{figure}
\epsscale{0.49} 
%\plotone{Ueta0r0rs.w.ps}
%\plotone{Ueta0r0rs.1.ps}
%\plotone{Ueta0r0rs.2.ps}
%\plotone{Ueta0r0rs.s.ps}
\plotone{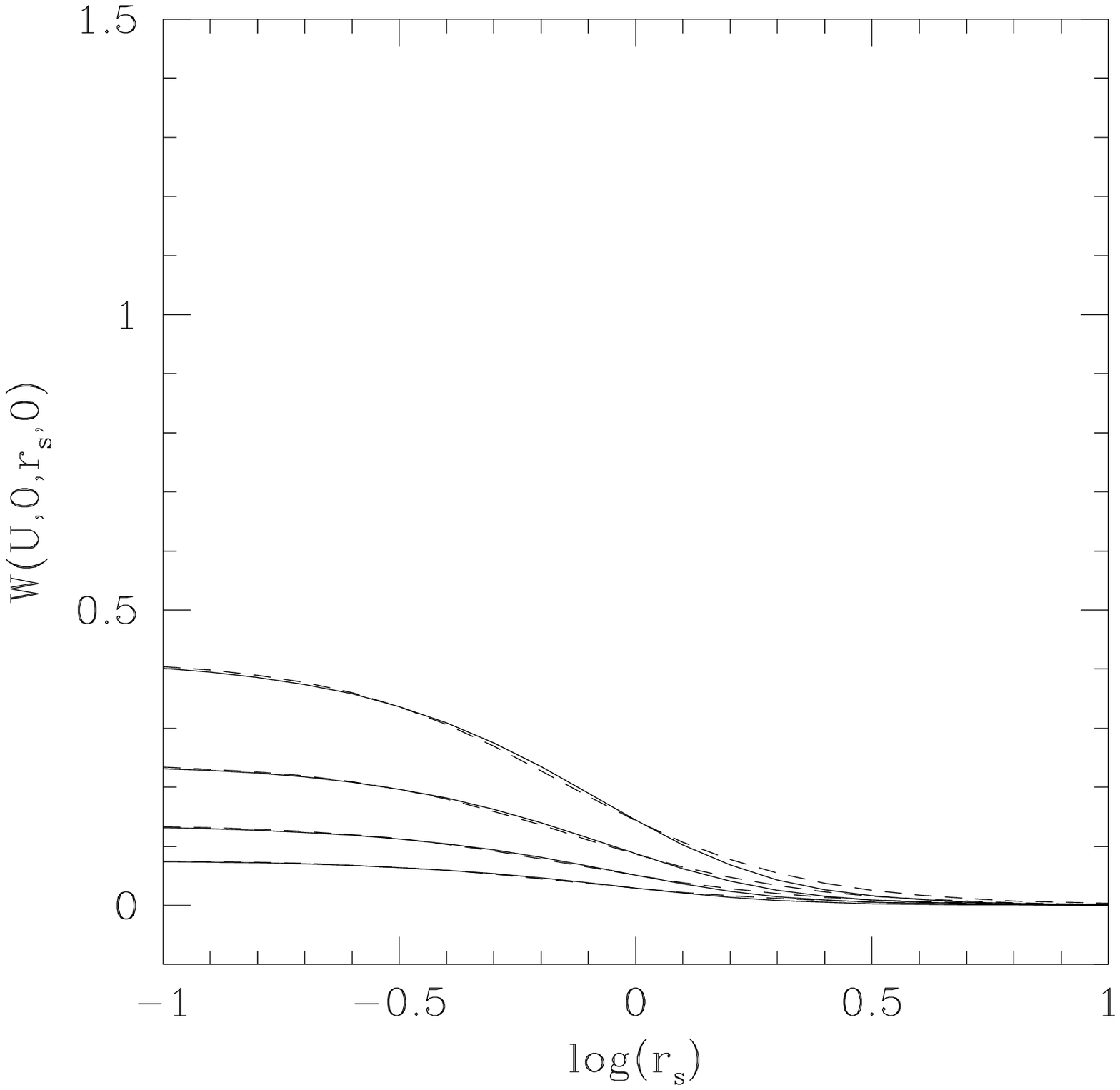}
\plotone{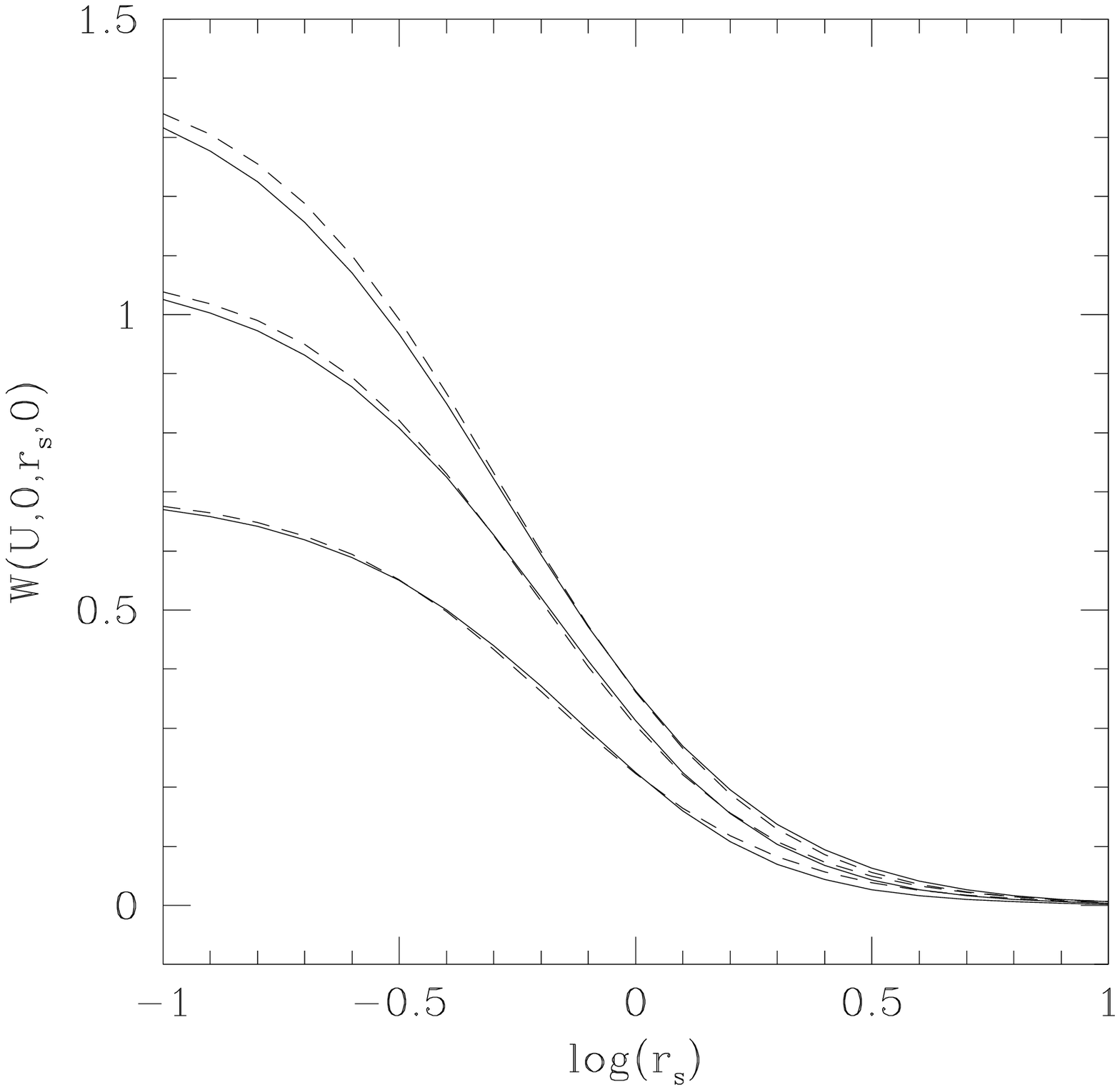}
\plotone{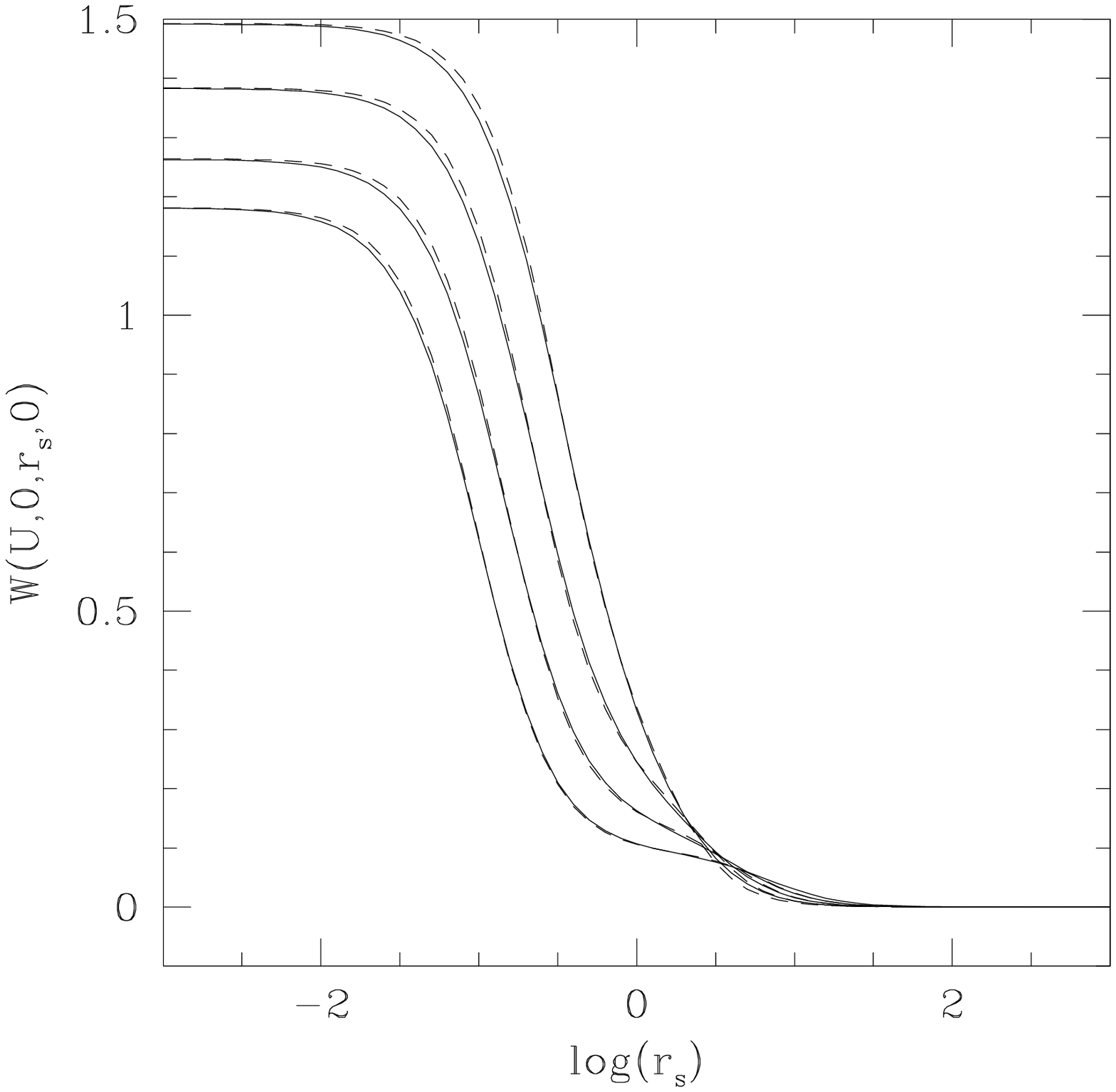}
\plotone{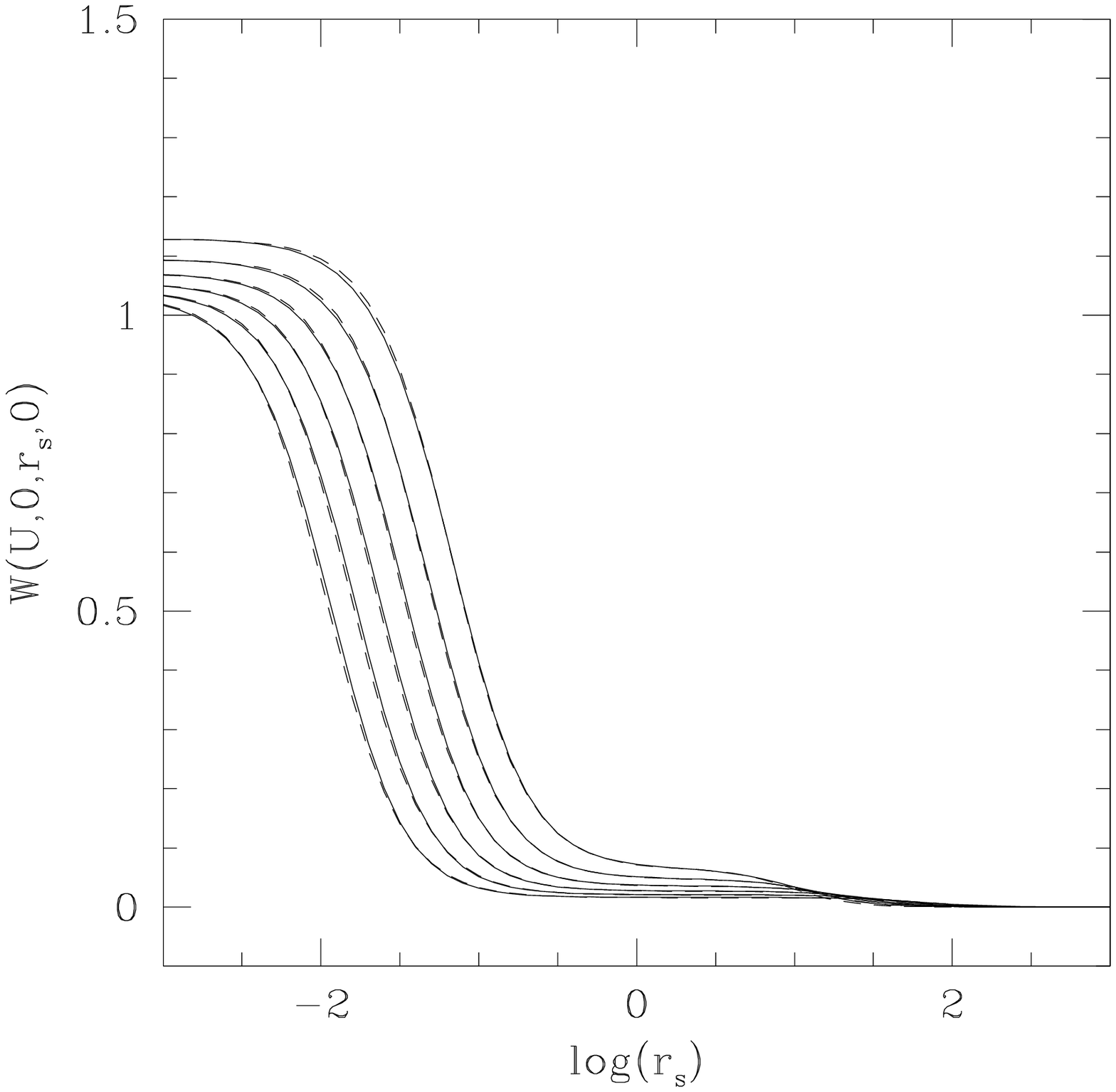}
\caption{Solid lines show the correlation as a function of source size
$r_s$ for zero separation ($r=0$) and zero wavelength difference ($\eta=0$)
as calculated numerically.  Dashed lines show the corresponding
fitting function.  Top Left: $\log(U) = -0.25, ~-0.5, ~-0.75, ~-1.0$
(from above).  Top Right: $\log(U) = 0.5, ~0.25, ~0$ (from above).
Bottom Left: $\log(U) = 0.75, ~1.0, ~1.25, ~1.5$ (from above).  Bottom
Right: $\log(U) = 1.75, ~2.0, ~2.25, ~2.5, ~2.75, ~3.0$ (from above).}
\label{figWrs}
\end{figure}

\begin{figure}
\epsscale{0.49} 
%\plotone{Uetar0rs0.w.ps}
%\plotone{Uetar0rs0.1.ps}
%\plotone{Uetar0rs0.2.ps}
%\plotone{Uetar0rs0.s.ps}
\plotone{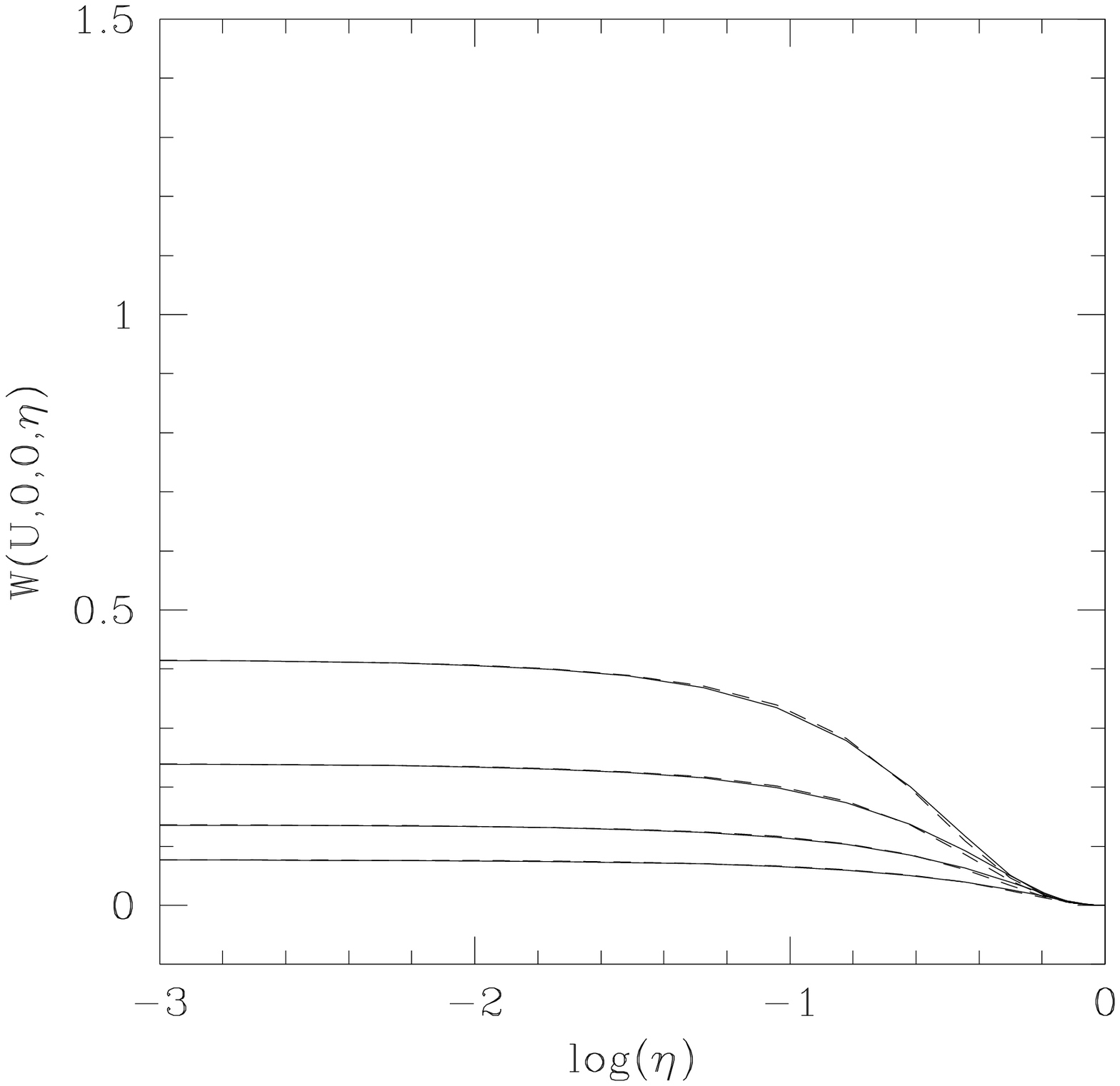}
\plotone{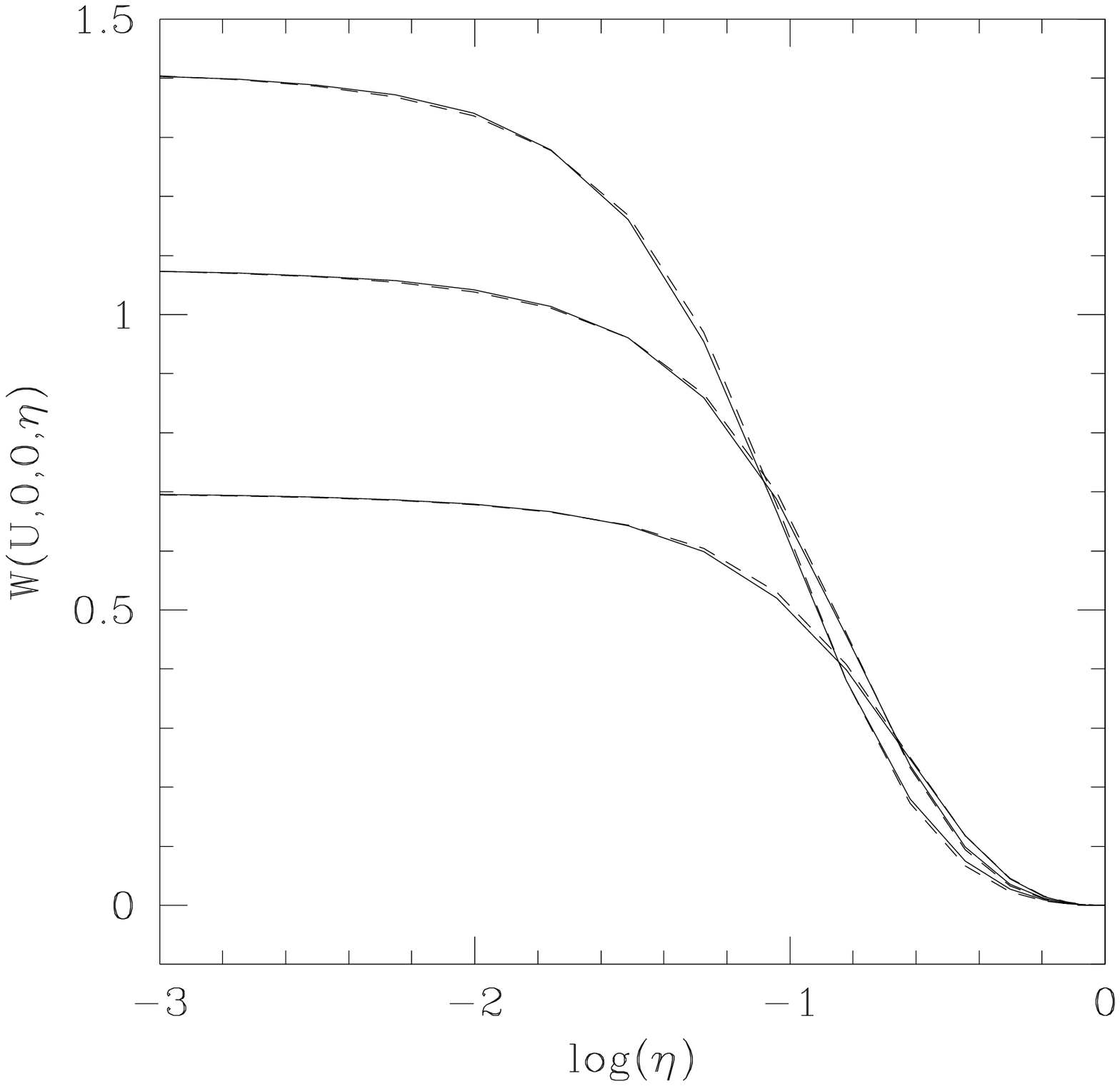}
\plotone{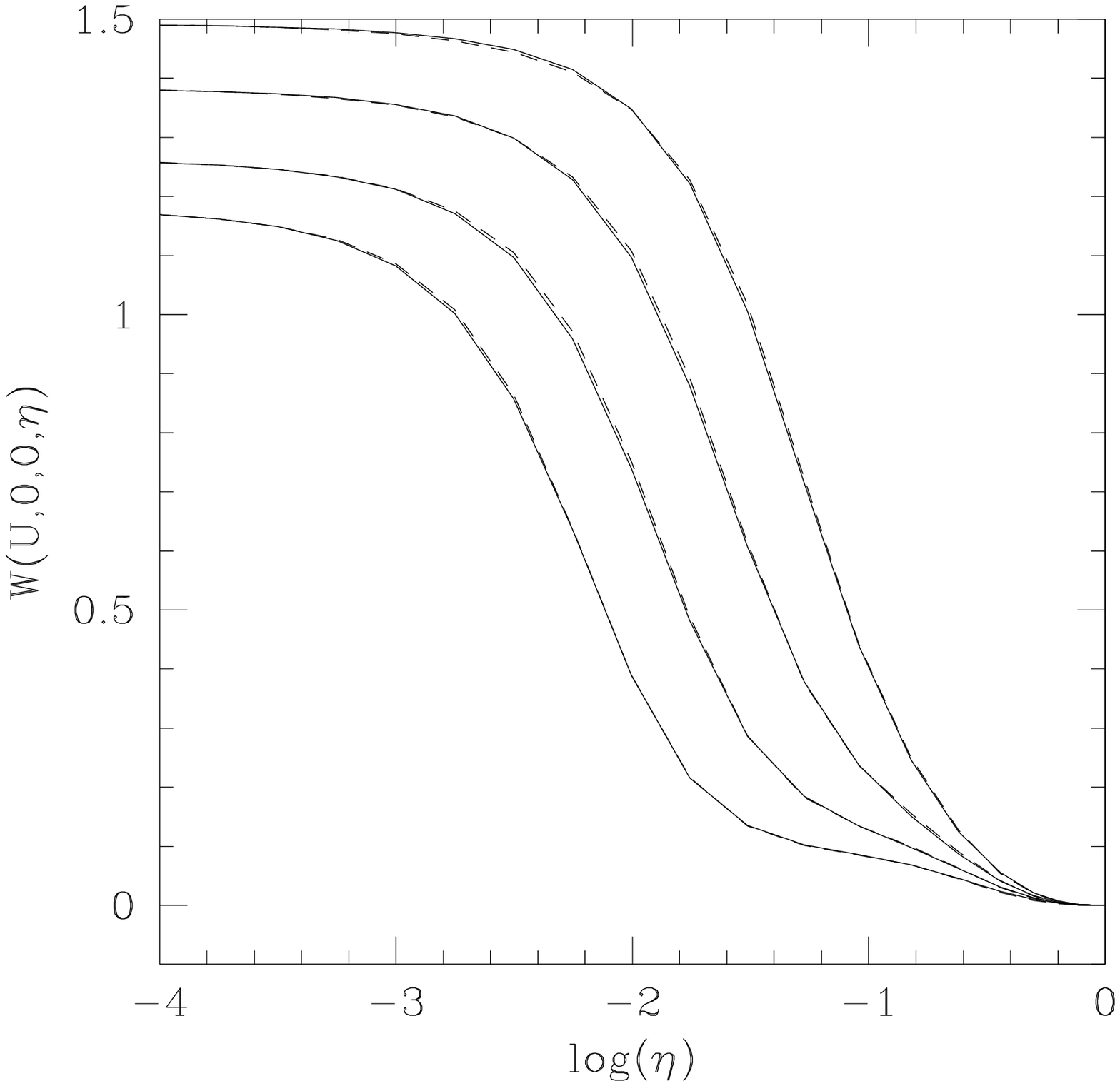}
\plotone{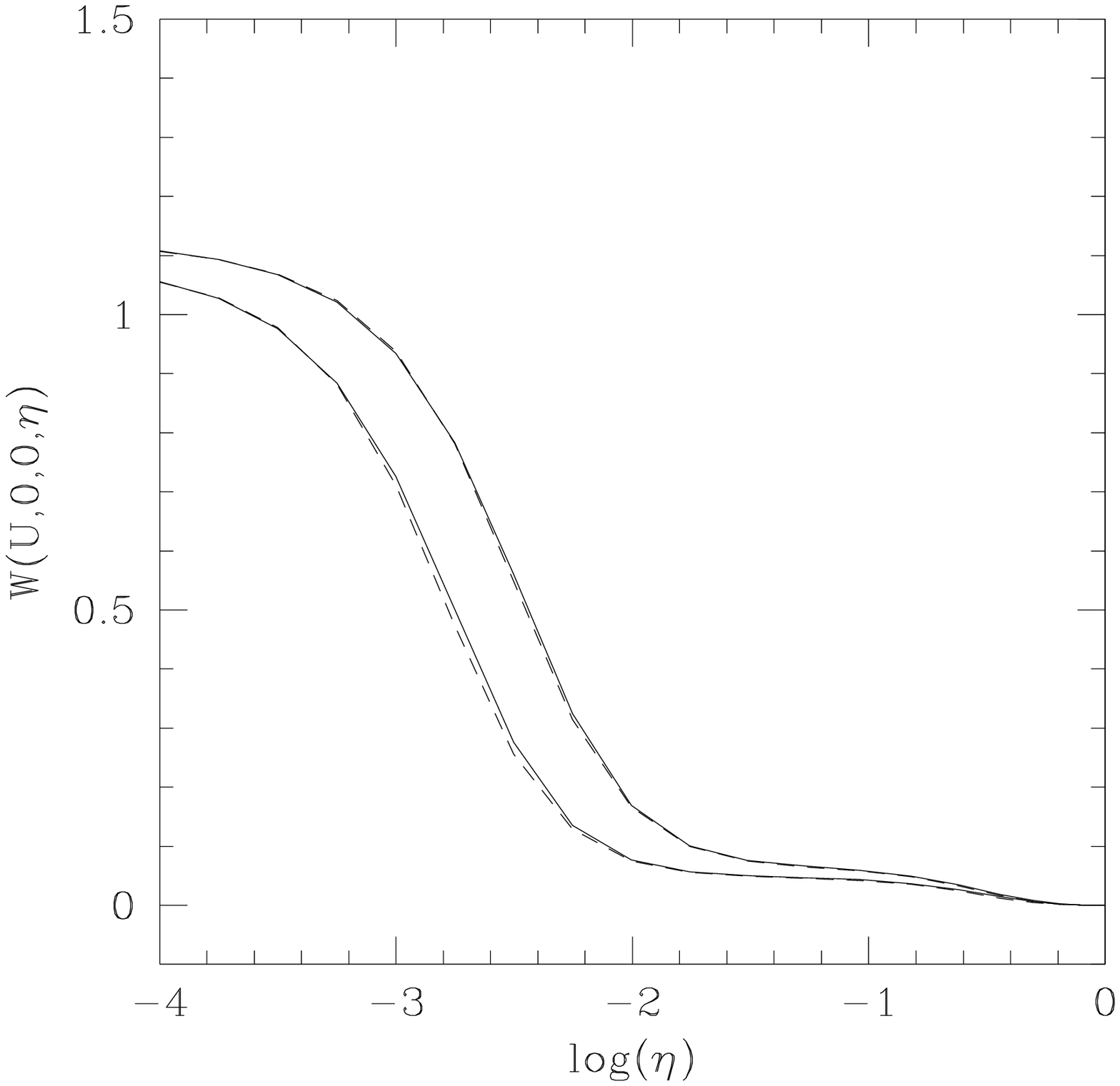}
\caption{Solid lines show the correlation as a function of wavelength
difference $\eta$ for zero separation ($r=0$) and zero source size ($r_s=0$)
as calculated numerically.  Dashed lines show the corresponding
fitting function.  Top Left: $\log(U) = -0.25, ~-0.5, ~-0.75, ~-1.0$
(from above).  Top Right: $\log(U) = 0.5, ~0.25, ~0$ (from above).
Bottom Left: $\log(U) = 0.75, ~1.0, ~1.25, ~1.5$ (from above).  Bottom
Right: $\log(U) = 1.75, ~2.0$ (from above).}
\label{figWeta}
\end{figure}

Figures \ref{figWr}, \ref{figWrs}, \ref{figWeta} show the dependence
of the flux correlation function $W(U,r,r_s,\eta)$ on each of the
three variables $r$, $r_s$ and $\eta$, while keeping the other two
variables fixed at zero.  Both the exact numerical results and the
fitting function are shown.

%\clearpage

\begin{figure}
\epsscale{0.49} 
%\plotone{U1eta0rrs.ps}
%\plotone{U3.162eta0rrs.ps}
%\plotone{U10eta0rrs.ps}
%\plotone{U31.62eta0rrs.ps}
\plotone{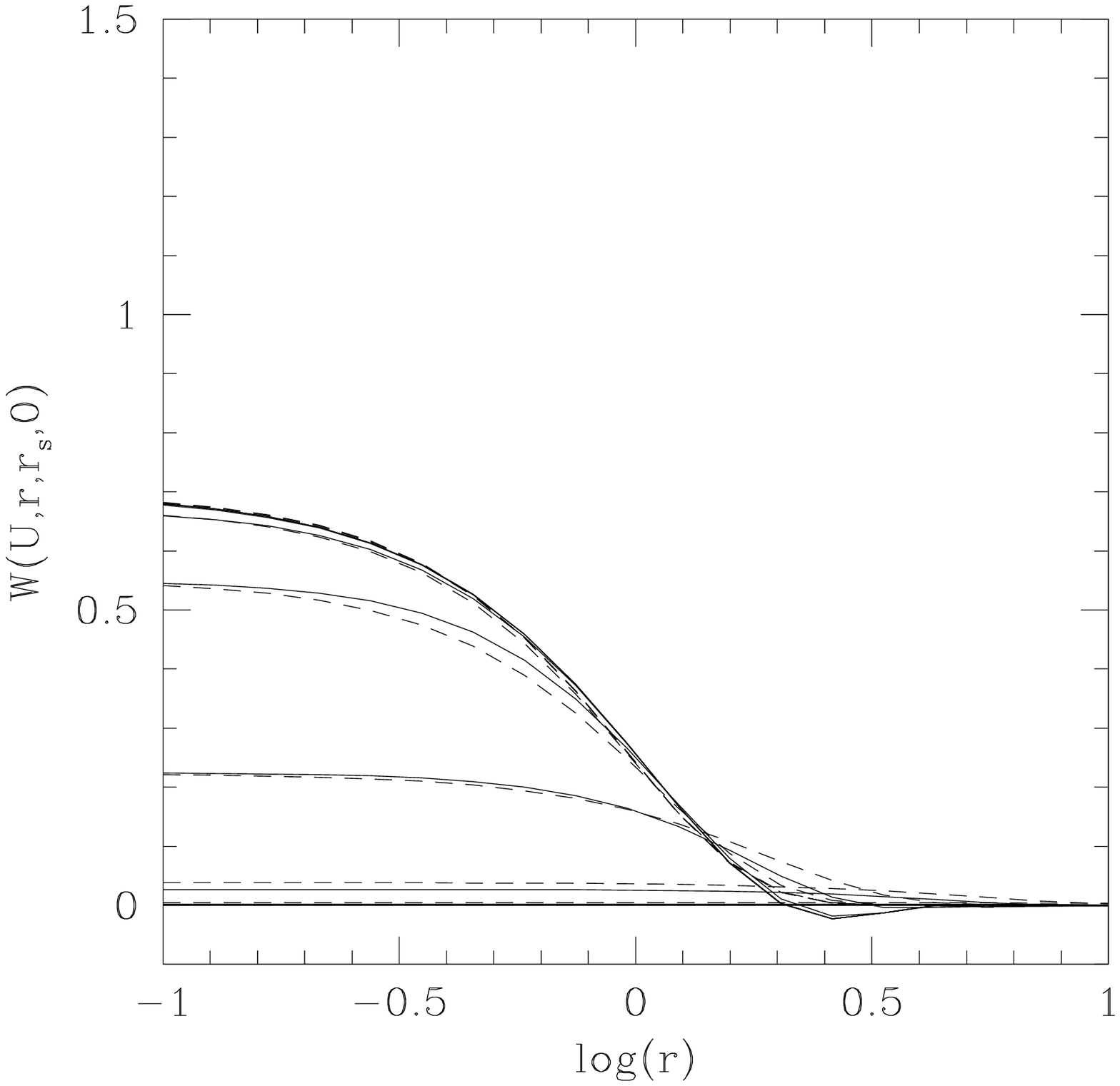}
\plotone{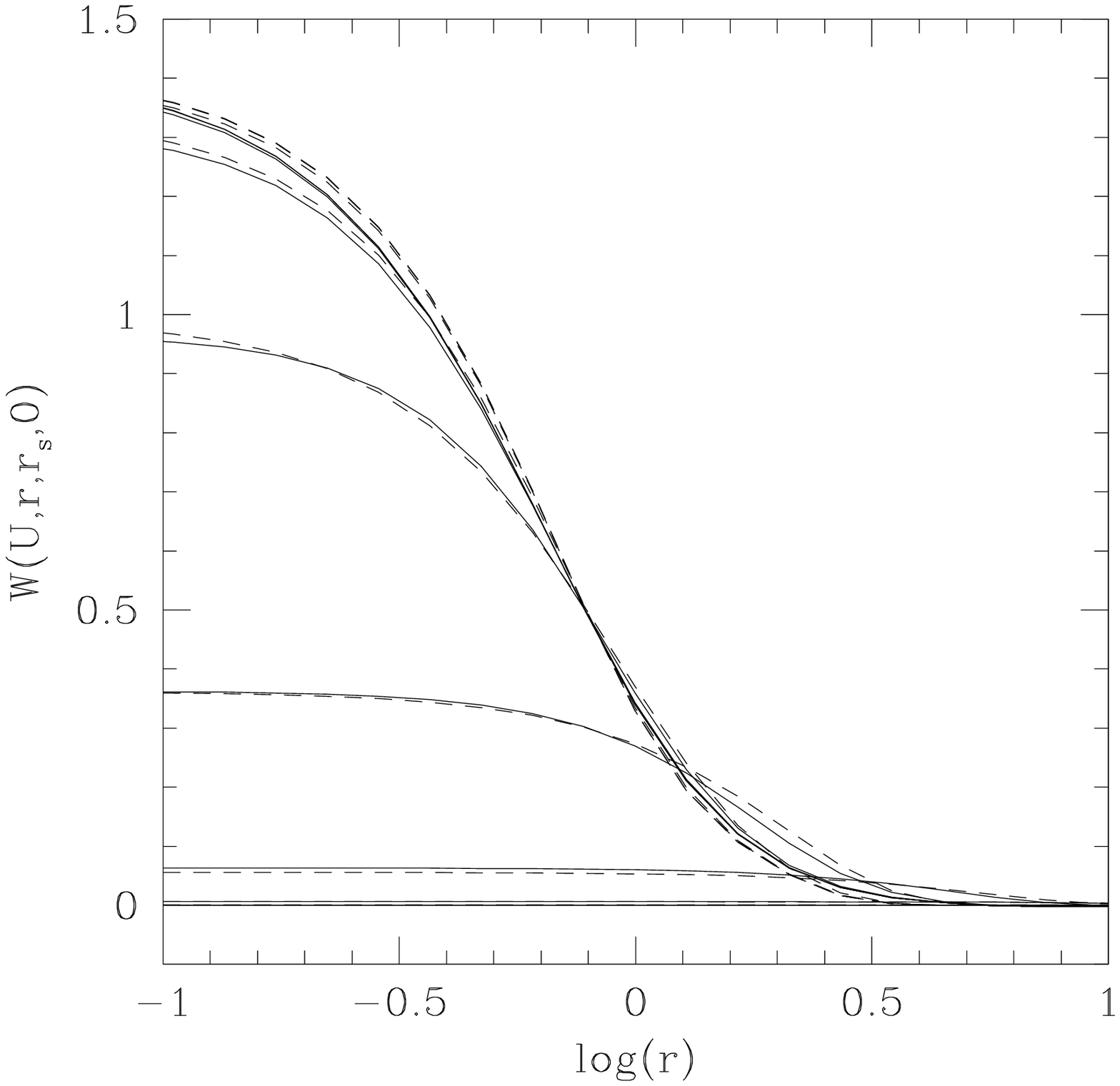}
\plotone{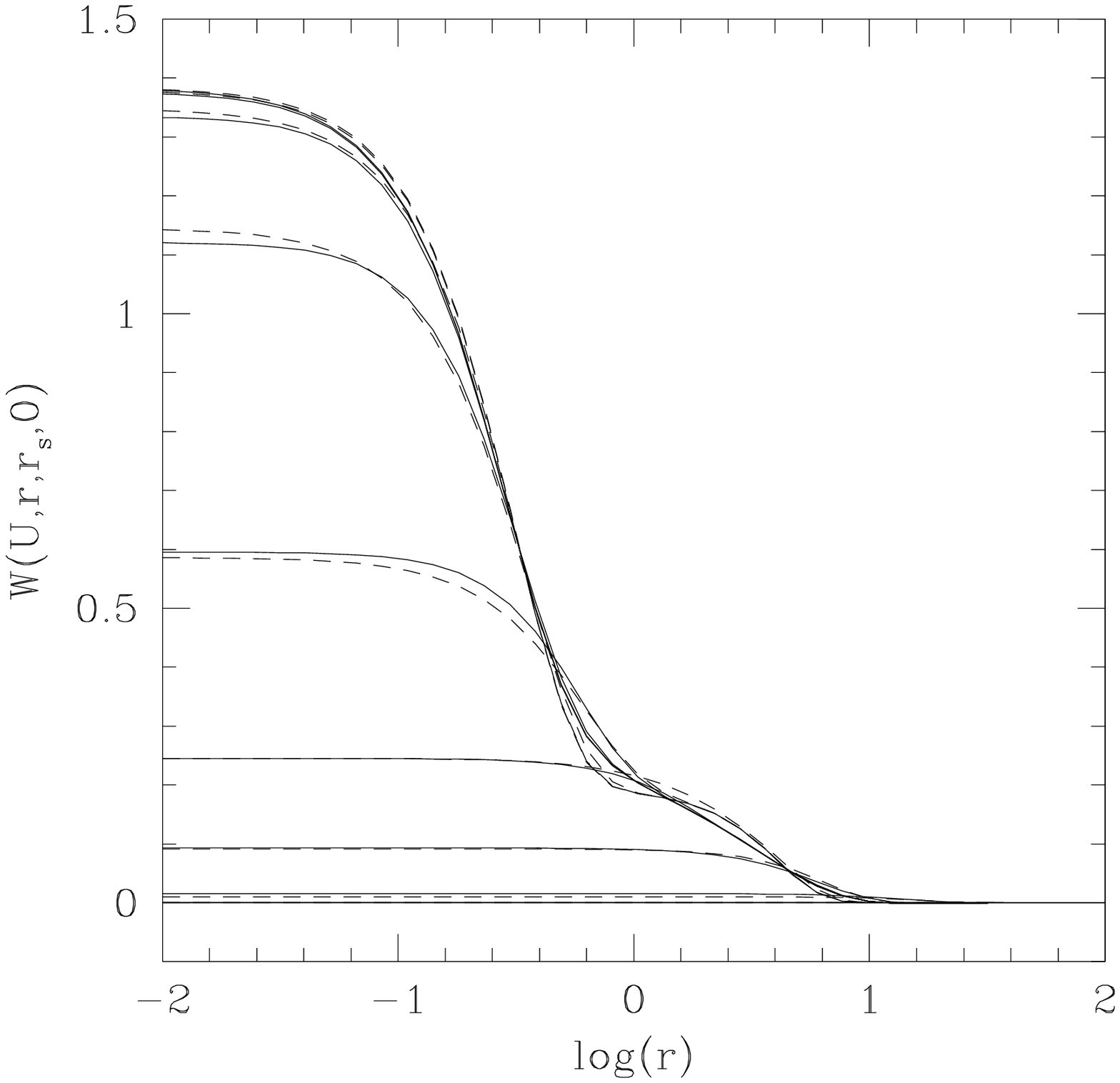}
\plotone{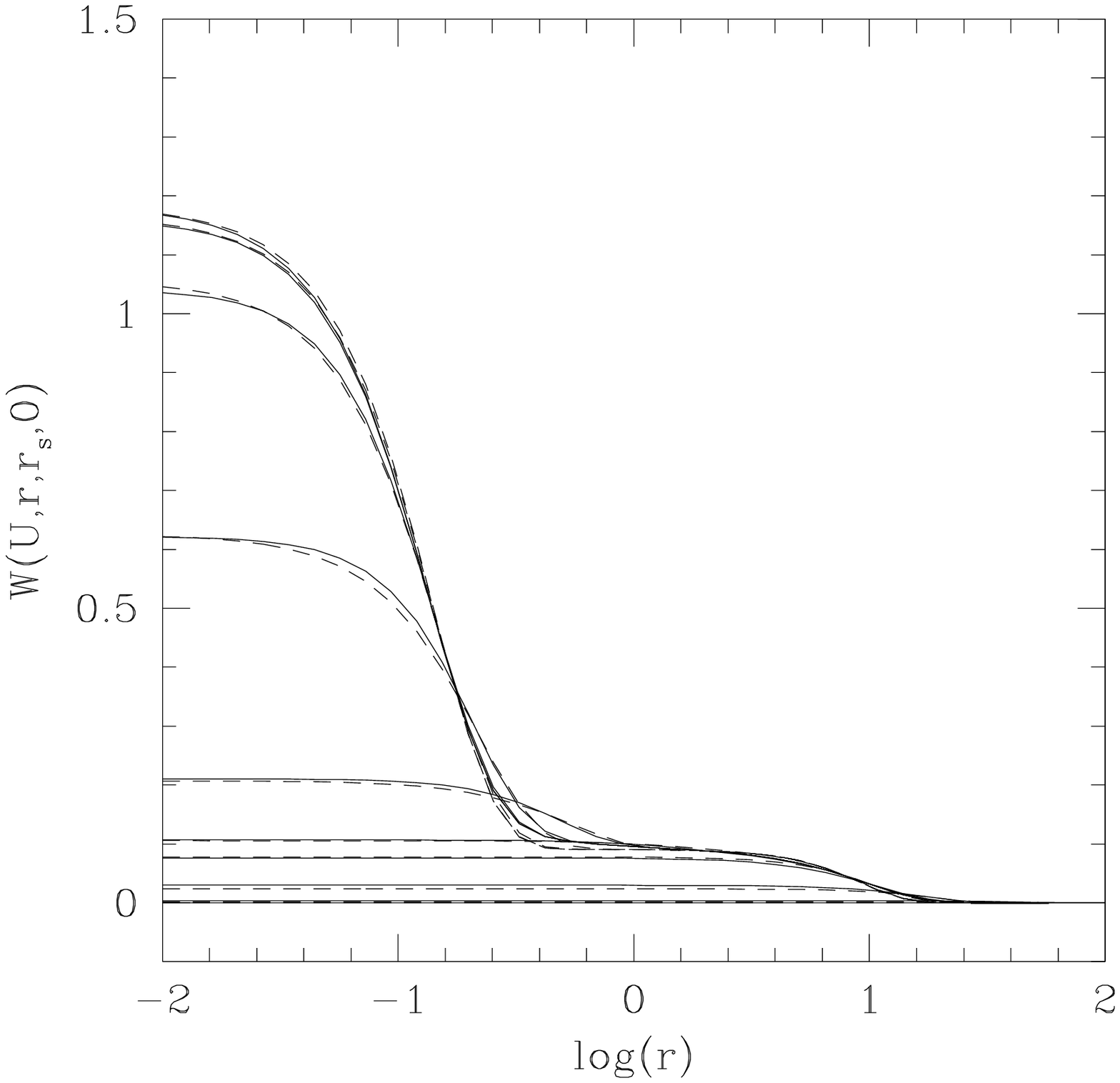}
\caption{Solid lines show the numerically calculated correlation as a function 
of separation $r$ for a series of values of $r_s$, for $\eta=0$.  In
each panel, from above, the curves correspond to $r_s = 0, ~10^{-2},
~10^{-1.5}, ~10^{-1}, ~10^{-0.5}, ~1, ~10^{0.5}, ~10^{1}, ~10^{1.5},
~10^{2}$.  The dashed lines show the fitting function.  Top Left:
$U=1$.  Top Right: $U=3.162$.  Bottom Left: $U=10$.  Bottom Right:
$U=31.62$.}
\label{figWrrs}
\end{figure}

\begin{figure}
\epsscale{0.49}
%\plotone{U1etarrs0.ps}
%\plotone{U3.162etarrs0.ps}
%\plotone{U10etarrs0.ps}
%\plotone{U31.62etarrs0.ps}
\plotone{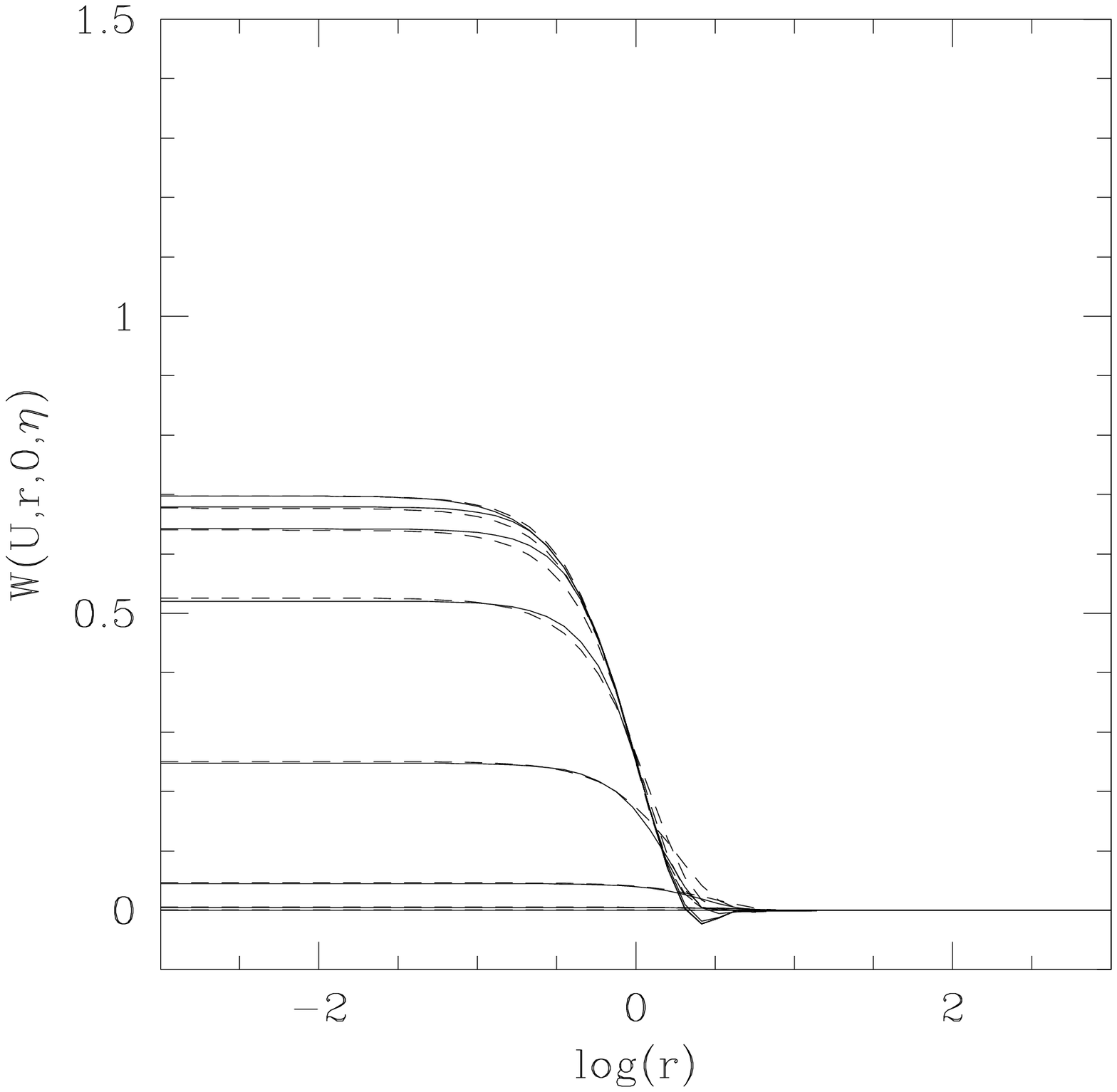}
\plotone{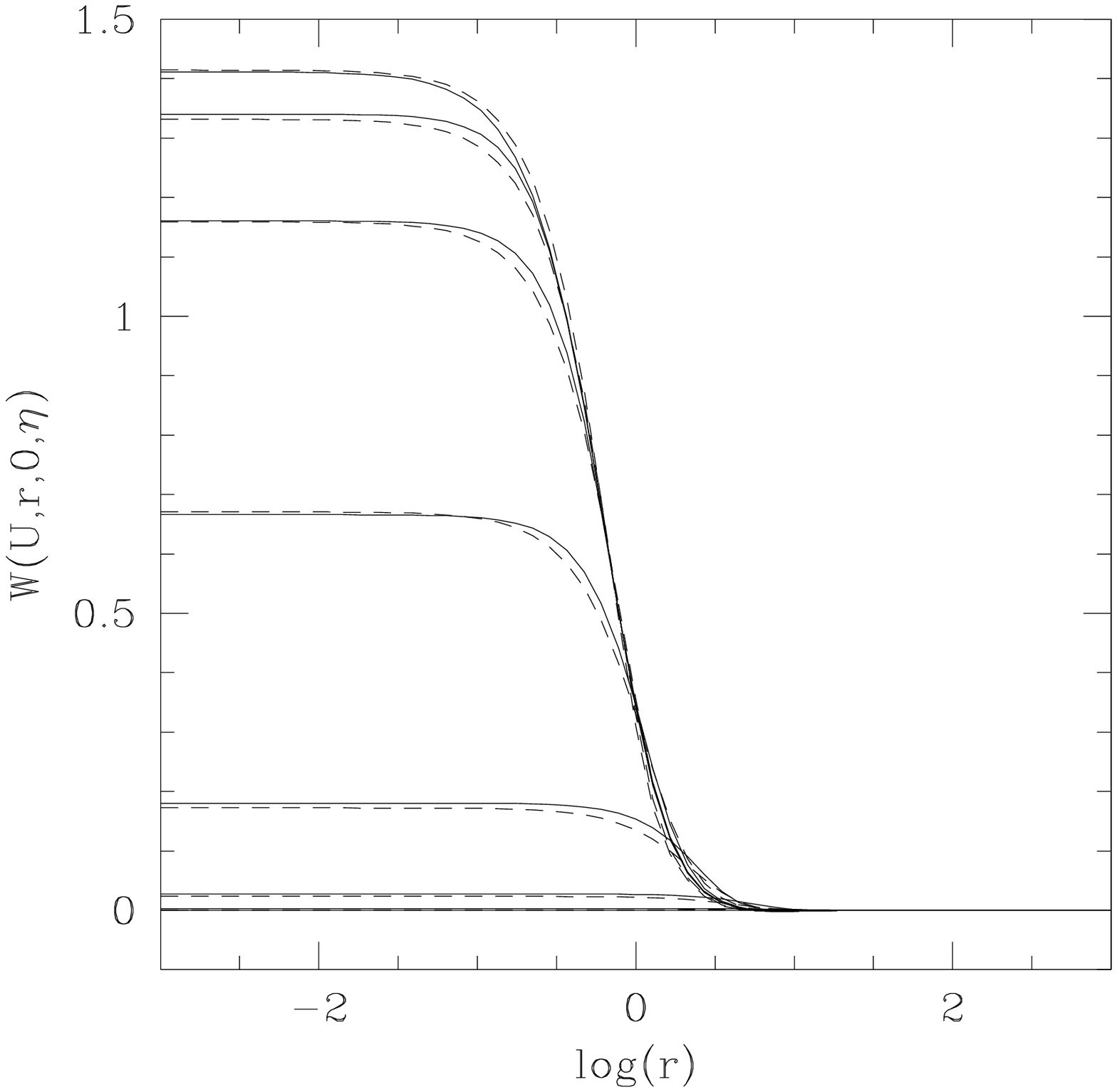}
\plotone{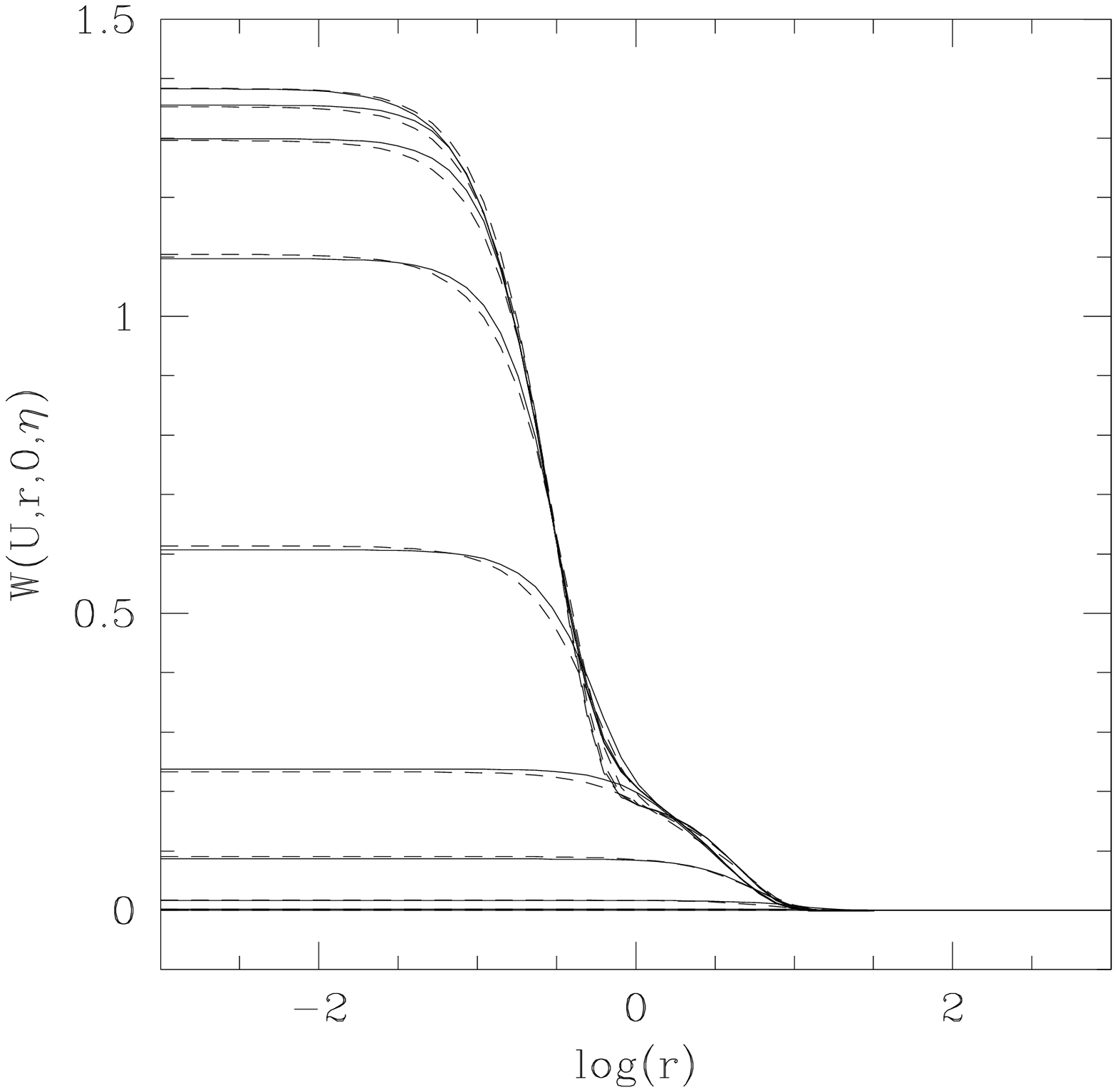}
\plotone{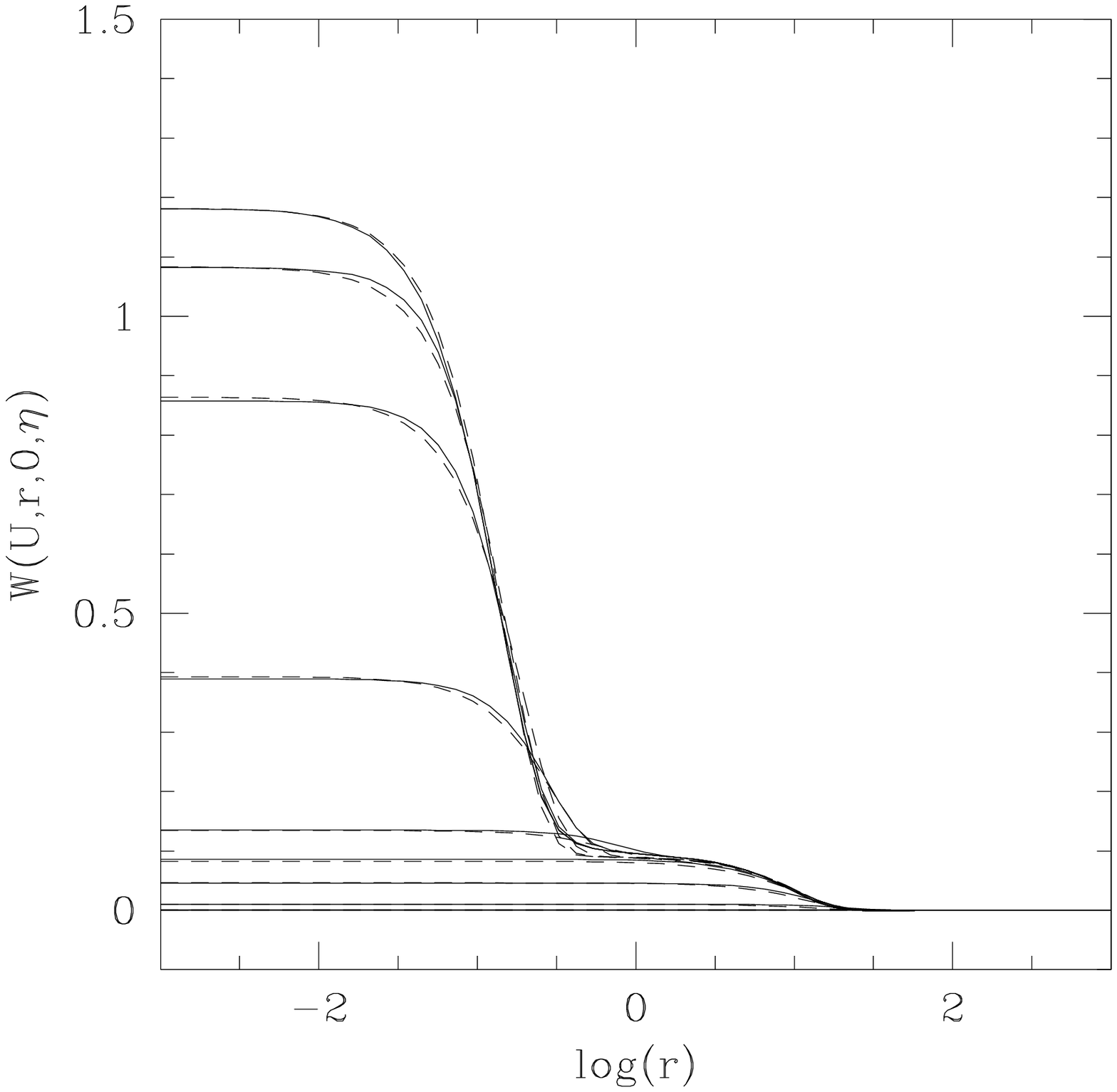}
\caption{Solid lines show the numerically calculated correlation as a 
function of separation $r$ for a series of values of $\zeta$ (see
eq. \ref{zetadef} for the definition), for $r_s=0$.  In each panel,
from above, the curves correspond to $\zeta = 0, ~10^{-2}, ~10^{-1.5},
~10^{-1}, ~10^{-0.5}, ~1, ~10^{0.5}, ~10$.  The dashed lines show the
fitting function.  Top Left: $U=1$.  Top Right: $U=3.162$.  Bottom
Left: $U=10$.  Bottom Right: $U=31.62$.}
\label{figWetar}
\end{figure}

Figure \ref{figWr} shows that the flux correlation decays with
increasing lag.  The decay is described by a single lengthscale when
$U$ is less than a few (weak scattering regime), whereas the decay
occurs on two distinct lengthscales for larger values of $U$ (strong
scattering).  In the latter regime, the shorter of the two scales is
the diffractive scale and the longer is the refractive scale (see
Rickett 1990; Narayan 1993; for a review of the underlying physics of
diffractive and refractive scintillation), and the ratio of the two
scales increases with inreasing $U$ as $U^{6/5}$.  The fluctuation
power on the diffractive scale remains large for all values of $U$,
whereas that on the refractive scale decays with increasing $U$.  All
of these properties are well-known and asymptotic results are
available in the limit of both very large and very small $U$.  The
fitting function we have developed handles the asymptotic regimes
well, but more importantly, it also models the analytically
intractable regime corresponding to $U\sim 1-10$ quite satisfactorily.
The most serious qualitative discrepancy is that the fitting function
gives a positive value of the correlation for all lags, whereas the
numerically computed correlation sometimes goes negative, especially
for $U\sim 1$ (e.g., the top two panels of
Fig. \ref{figWr}; see also Figs. \ref{figWrrs}, \ref{figWetar}).
More pronounced negative ``overshoot'' is seen in scintillation of
some extragalactic intra-day variables and has been interpreted
as evidence for anisotropic scattering\citep{ric02,big03}.

Figure \ref{figWrs} shows the dependence of the flux correlation as a
function of the source size $r_s$, and Figure \ref{figWeta} shows the
dependence of the cross-correlation as a function of the dimensionless
frequency difference $\eta$.  Once again, the results clearly indicate
the transition from a single scale in weak scattering to two scales in
strong scattering.  The fitting function does a remarkably good job
for all values of $U$ and all choices of $r_s$ and $\eta$.

\begin{figure}
\epsscale{0.49}
%\plotone{U1etar0rs.ps}
%\plotone{U3.162etar0rs.ps}
%\plotone{U10etar0rs.ps}
%\plotone{U31.62etar0rs.ps}
\plotone{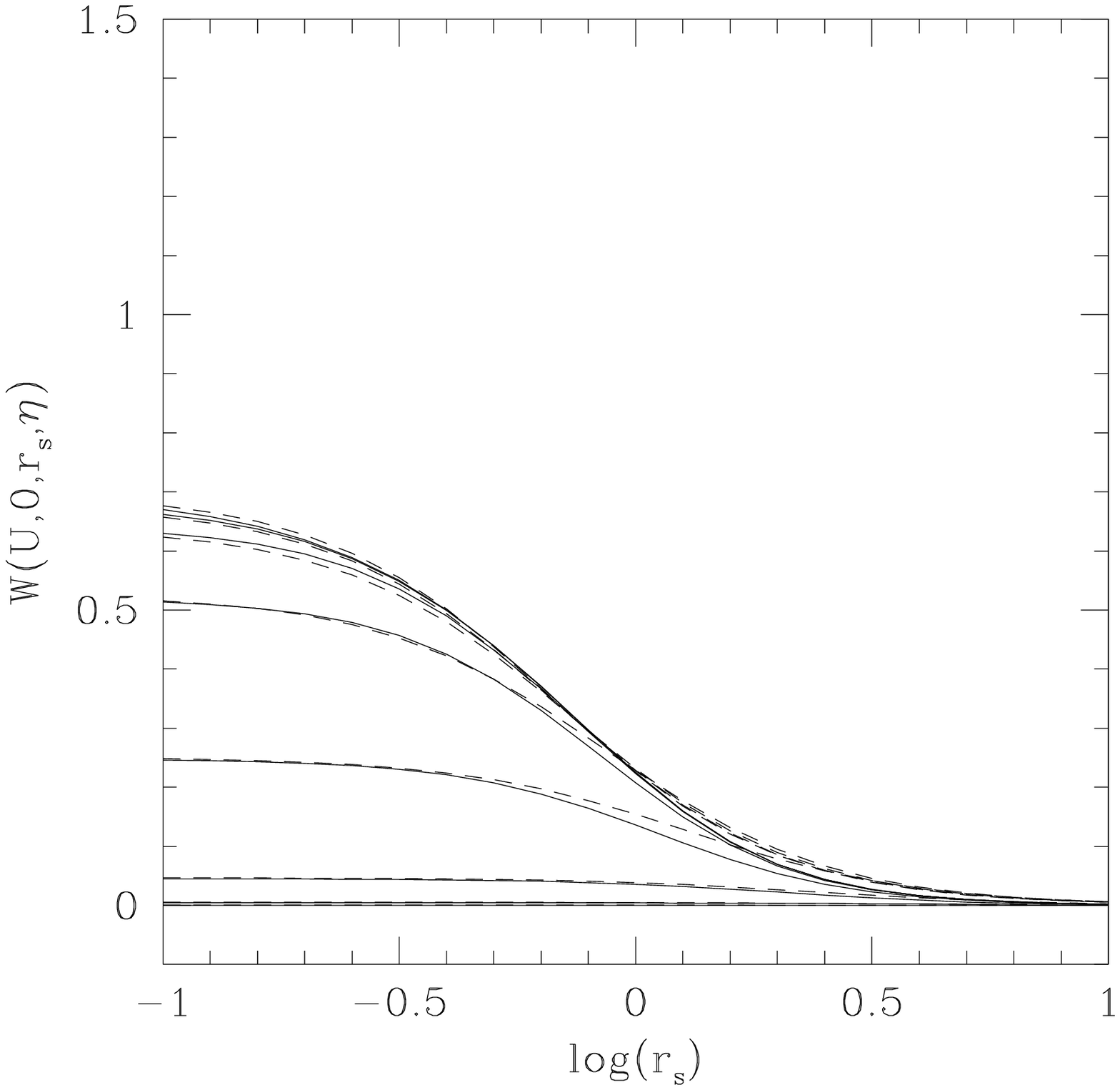}
\plotone{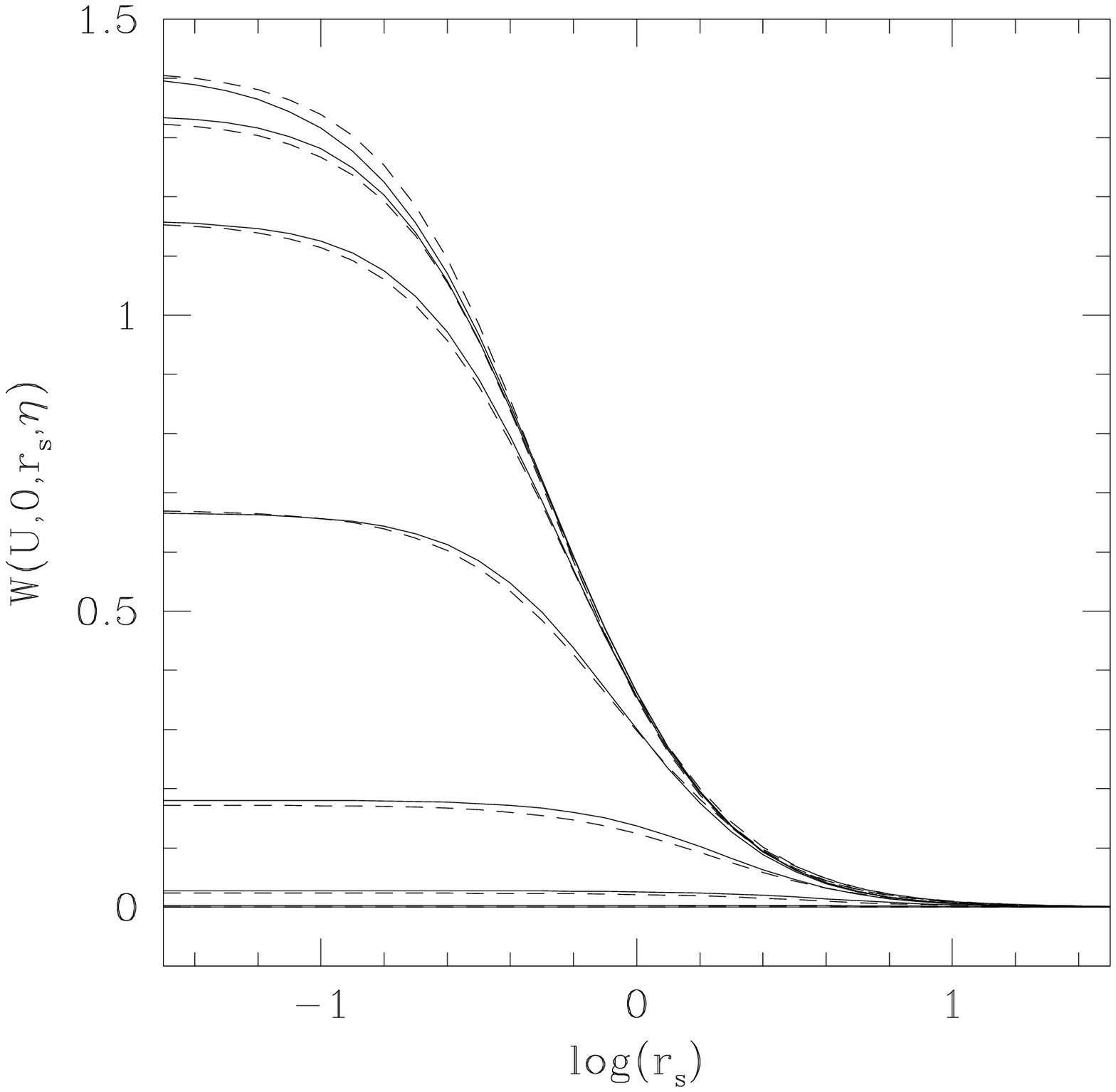}
\plotone{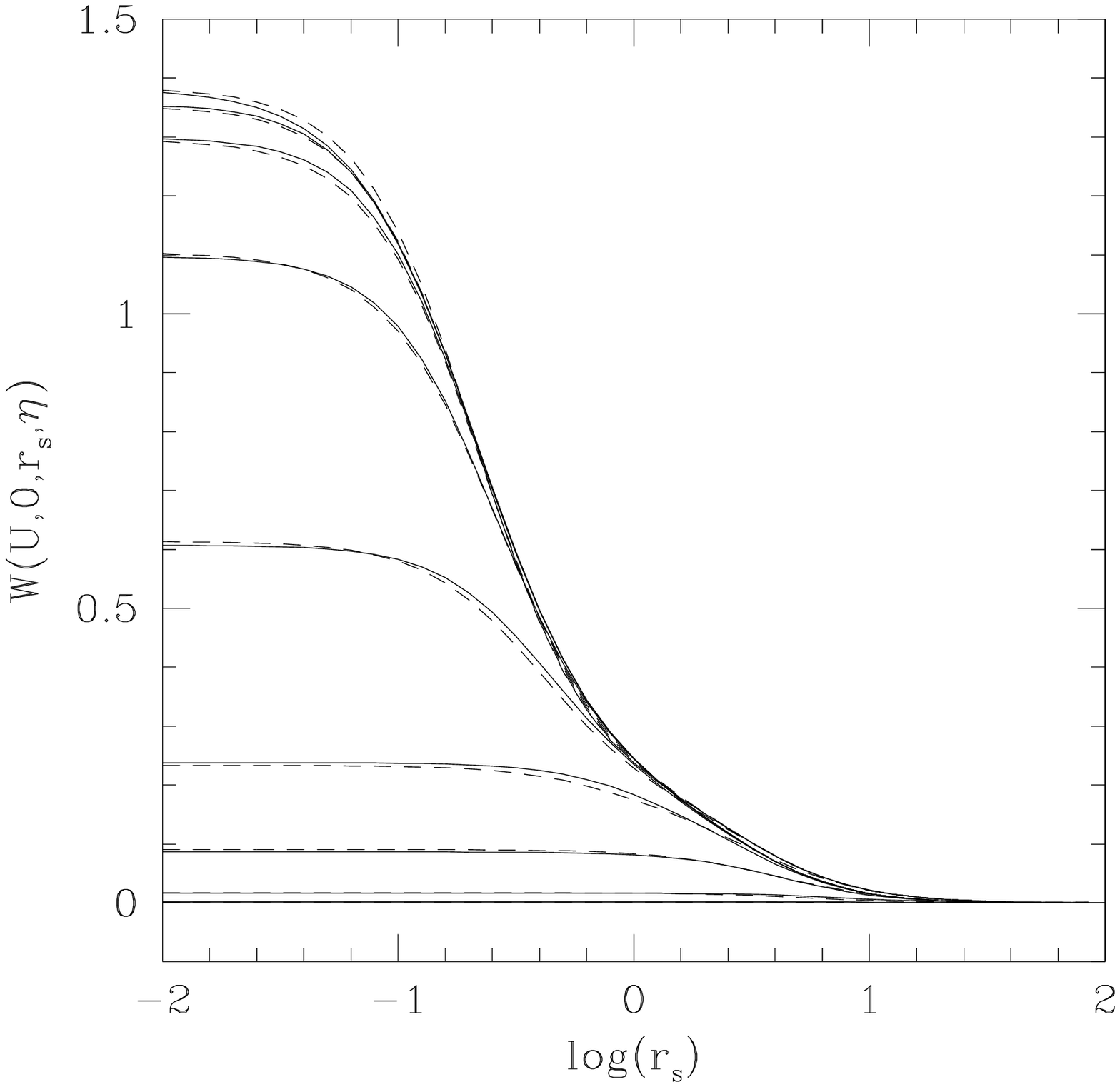}
\plotone{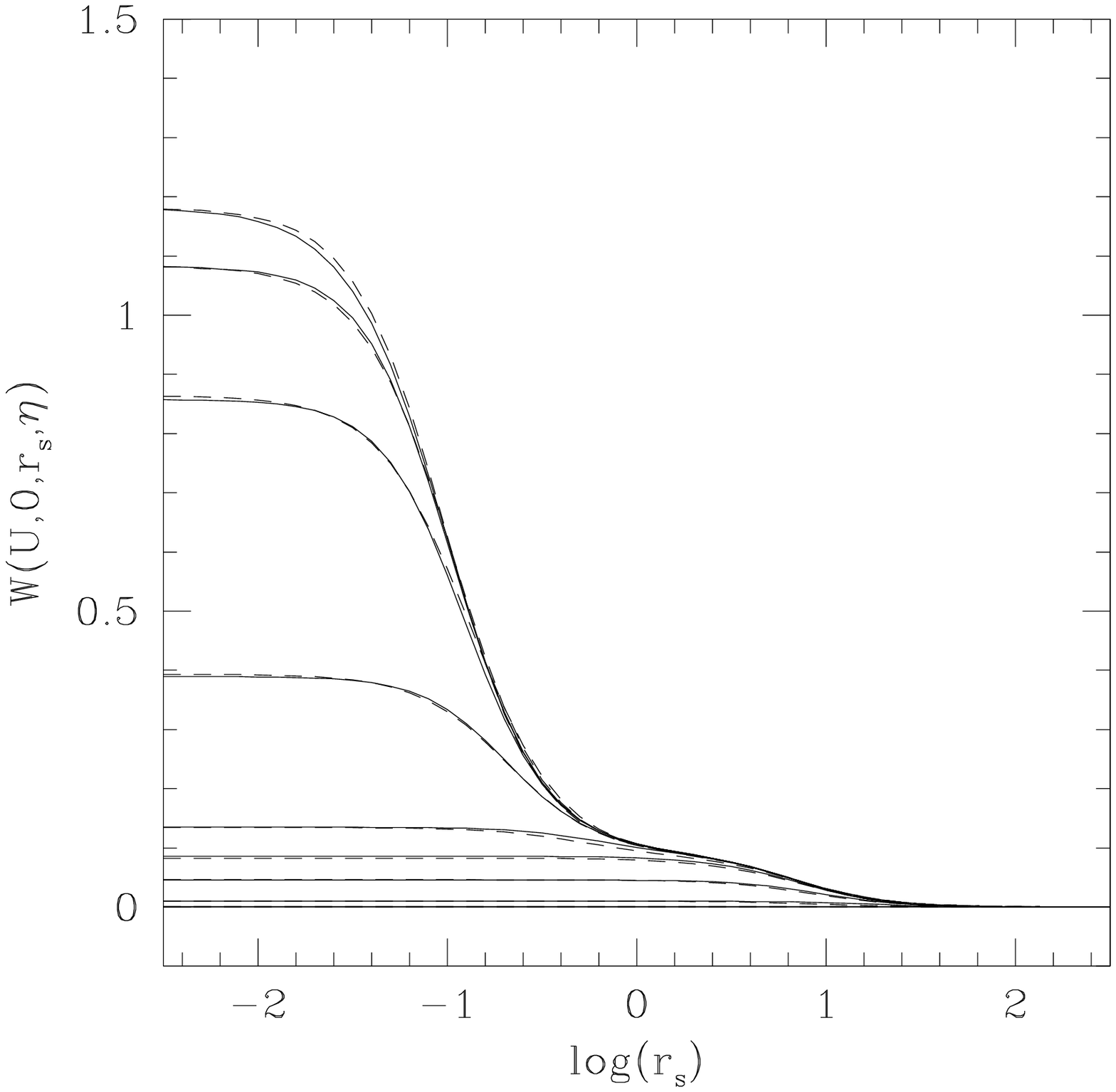}
\caption{Solid lines show the correlation as a function of source size
$r_s$ for a series of values of $\zeta$, for $r=0$.  In each panel,
from above, the curves correspond to $\zeta = 0, ~10^{-2}, ~10^{-1.5},
~10^{-1}, ~10^{-0.5}, ~1, ~10^{0.5}, ~10$.  The dashed lines show the
fitting function.  Top Left: $U=1$.  Top Right: $U=3.162$.  Bottom
Left: $U=10$.  Bottom Right: $U=31.62$.}
\label{figWetars}
\end{figure}

Figures \ref{figWrrs}, \ref{figWetar}, \ref{figWetars} show the
dependence of $W(U,r,r_s,\eta)$ as a function of pairs of the three
variables, $r$, $r_s$, $\eta$, keeping the third variable fixed at
zero.  These two-dimensional correlations are relatively less well
explored in the literature, even in the asymptotic regimes, but our
results are generally consistent with previous work wherever a
comparison is possible.  The behavior of $W(U,r,r_s,0)$ in Figure
\ref{figWrrs} is easy to understand.  With increasing source size, the
fluctuations are progressively smoothed, so the fluctuation amplitude
decreases and the scale of the fluctuations increases, just as one
would expect.  The dependences in Figures \ref{figWetar} and
\ref{figWetars} are less obvious.  With increasing frequency
difference $\eta$, the decay scale of the correlation increases both
as a function of $r$ (Fig. \ref{figWetar}) and as a function of $r_s$
(Fig.  \ref{figWetars}); this is especially obvious in the plots for
weak scintillation and strong diffractive scintillation.
Equivalently, the decorrelation bandwidth increases with increasing
$r$ and/or $r_s$.  Chashei \& Shishov (1976) showed such a dependence
for strong diffractive scintillation for the particular case of a
``square law'' spectrum, $\alpha=2$.  No analytical results have been
reported for a Kolmogorov spectrum $\alpha=5/3$, but we see from our
numerical work that the overall behavior is similar.  In any case,
Figures \ref{figWrrs}, \ref{figWetar} and \ref{figWetars} show that
the fitting function reproduces the exact numerical results quite well
for all values of $U$ and choices of $r$, $r_s$, $\eta$.

The behavior of flux scintillation in the $r-\eta$ plane has gained
prominence in recent years as a result of the discovery of parabolic
``arcs'' and ``arclets'' in the so-called secondary spectrum
\citep{stinebring01,cor04}.  Appendix C develops a
rudimentary analytical theory of arcs that is valid in the two
asymptotic regimes $U\ll1$ and $U\gg1$.  The discussion there
complements the more detailed physical approach taken by Cordes et
al. (2005).  The behavior of arcs in the transition regime $U\sim$
few, and in the presence of anisotropic turbulence, deserves further
analytical and numerical study, but it is beyond the scope of this
work.  Researchers interested in arcs should be warned that our fitting
functions were not designed with arcs in mind.

%\clearpage
%\begin{figure}
%\epsscale{1.1}
%\plotone{U1.0etarrs.ps}
%\caption{Solid lines show the correlation as a function of separation
%$r$ for a series of values of $r_s$.  In each panel, from above, the
%curves correspond to $r_s = 0, ~10^{-2}, ~10^{-1.5}, ~10^{-1},
%~10^{-0.5}, ~1, ~10^{0.5}, ~10^{1}, ~10^{1.5}, ~10^2$.  The dashed
%lines show the fitting function.  All panels have $U=1$ but different
%values of $\zeta$.  Top Left: $\zeta=0.03162$.  Top Right:
%$\zeta=0.1$.  Bottom Left: $\zeta=0.3162$.  Bottom Right: $\zeta=1$.}
%\end{figure}

%\begin{figure}
%\epsscale{1.1}
%\plotone{U10.0etarrs.ps}
%\caption{Solid lines show the correlation as a function of separation
%$r$ for a series of values of $r_s$.  In each panel, from above, the
%curves correspond to $r_s = 0, ~10^{-2}, ~10^{-1.5}, ~10^{-1},
%~10^{-0.5}, ~1, ~10^{0.5}, ~10^{1}, ~10^{1.5}, ~10^2$.  The dashed
%lines show the fitting function.  All panels have $U=10$ but different
%values of $\zeta$.  Top Left: $\zeta=0.001$.  Top Right: $\zeta=0.01$.
%Bottom Left: $\zeta=0.1$.  Bottom Right: $\zeta=1$.}
%\end{figure}

We have also compared the fitting function with the numerical results
on $W(U,r,r_s,\eta)$ for the case when all three variables, $r$,
$r_s$, $\eta$, are varied, but we do not show the corresponding plots.
By comparing the fitting function and the numerical results over an
extensive grid of values of $U$, $r$, $r_s$ and $\eta$, we find that
the maximum error in this four-dimensional space for a gaussian source
is 0.047.  The maximum error is larger for a tophat source,
$\sim0.06$, perhaps because the simplification of replacing the two
source sizes $r_{s1}$ and $r_{s2}$ with a single effective size $r_s$
(eq. \ref{rsdef}) is less well motivated in that case.

\subsection{Logic Behind the Fitting Function}

Although the fitting function described in the previous subsections
was obtained to a large extent by a combination of intuition and
trial-and-error, we tried to draw on analytical clues from Appendix B
wherever possible.

Consider first $W_0(U)$.  Equation (\ref{varweak}) shows that in the
limit of very weak scattering ($U \ll 1$) the flux variations have a
mean square amplitude equal to $0.7729U$; the first term in the
fitting function (\ref{W0fit}) is designed to satisfy this limit.
Similarly, for very strong scattering ($U \gg 1$), equation
(\ref{Wref0}) shows that the refractive fluctuations have an amplitude
$0.2380 U^{-0.4}$ while equation (\ref{diffvar}) indicates that the
diffractive fluctuations have amplitude $1 + 0.2380 U^{-0.4}$.  Thus,
the total mean square flux variations is equal to $1 + 0.4760
U^{-0.4}$, which is ensured by the second term in equation
(\ref{W0fit}) when $U\gg1$.  Also, we see that the diffractive
fluctuations have a baseline amplitude of unity and that the excess
fluctuations above unity are divided equally between diffractive and
refractive fluctuations.  This is the motivatation behind the
particular split used in equation (\ref{Wdrdef}).  The particular
functional form that we have chosen in equation (\ref{W0fit}) to
interpolate between the weak and strong scattering limits is
completely arbitrary.  In fact, it is easy to find other forms that
model the transition region around $U\sim1$ equally well.

Consider next the flux correlation as a function of separation $r$,
for a point source ($r_s=0$) and zero wavelength difference
($\eta=0$).  Equation (\ref{Wdiff}) shows that, for diffractive
scintillation in the strong scattering regime with $\alpha=5/3$, the
correlation varies as $\exp(-Ur^{5/3})$.  When $r_s=\eta=0$ and
$U\gg1$, the parameter $R_1$ in equation (\ref{Frdef}) tends to $R_5$
in equation (\ref{R5strong}), whose leading term is proportional to
$U^{-3/5}$.  Thus, the $r$-dependence of the fitting function has the
correct functional form.  Unfortunately, the analytical result
(\ref{Wweak}) does not give a very convenient expression for $F_r$ in
the weak scattering limit.  However, since we require the fitting
function to vary smoothly across the transition from strong
diffractive to weak scattering, we use the same functional form
(\ref{Frdef}) for both regimes.  Similarly, the expression
(\ref{Wref}) does not provide a useful approximation for the factor
$F_{r,r}$ for strong refractive scintillation.  By numerical
experimentation we determined that $F_{r,r}$ cuts off more rapidly
compared to $F_r$; we chose the index in (\ref{Frrdef}) to be 7/3 out
of a sense of ``symmetry.''

Appendix \ref{subsec:finitesource} discusses some asymptotic results
for a finite source size.  For large source size, the flux correlation
is shown to decline as a power law in $r_s$, though with a different
index in the different regimes.  In refractive scintillation, the
decline is described by $W \propto (r_s/r_{\rm ref})^{-7/3}$
(eq. \ref{Wreffinitesource} for $\alpha=5/3$) and this is hardwired
into the fitting function via equation (\ref{Fsrdef}).  The situation
in the case of weak and strong diffractive scintillation is more
complicated.  The former has a scaling $W\propto r_s^{-7/3}$
(eq. \ref{Wweakfinitesource}) and the latter $W\propto (r_s/r_{\rm
diff})^{-\beta}$ with $\beta<2$ (eq. \ref{Wdifffinitesource}).  The
two indices are thus different.  However, we have required our fitting
function to have a common set of scalings for weak and diffractive
scintillation.  (This is in the interests of simplicity and for
smoothness across $U=U_0$).  Because of this, we allowed the index on
$(r_s/R_2)$ in equation (\ref{Fsdef}) to be a free parameter and
adjusted it to give the best overall fit to the numerical data; this
resulted in a value for the index of 1.81.

Finally, consider the correlation as a function of the parameter
$\eta$ for $r=r_s=0$.  First, notice that $\eta$ is defined such that
it is equal to $\Delta\lambda/\lambda$ when the two wavelengths are
close to each other, but it tends to unity as the wavelengths differ
by an arbitrarily large amount.  Because of the latter property, we
have found it more convenient to use the modified parameter $\zeta$
defined in equation (\ref{zetadef}) in the fitting function.  This
parameter is equal to $\eta$ when $\eta \ll 1$, but it is proportional
to the wavelength ratio $\lambda_2/\lambda_1$ as $\eta \to 1$.

In weak scattering and when $\eta, ~\zeta \ll 1$, equation
(\ref{varweak}) shows that the correlation varies as $1-\eta^{5/6}$,
i.e., roughly as $(1+\zeta^{5/6})^{-1}$.  The functional form chosen
for $F_\eta$ in equation (\ref{Fetadef}) is designed to satisfy this
limiting result (note that $a_1$ tends to a constant when $U\ll1$).
In the strong diffractive regime and in the limit of large $\eta$
(large compared to the decorrelation bandwidth but still small
compared to unity), equation (\ref{largeV}) shows that the correlation
declines as $\eta^{-2}U^{-2.4}$.  This asymptotic dependence is
ensured by the term $a_2(U)\zeta^2$ in equation (\ref{Fetadef}).
Finally, in the strong refractive regime, equation (\ref{Wref0}) shows
that the correlation varies as $1-\eta^2$ for small $\eta$.  This is
taken care of by the term $a_4(U)\zeta^2$ in equation
(\ref{Fetardef}), where we note that $a_4$ tends to a constant in the
limit when $U\gg1$.

While we have tried to respect asymptotic results for $U\ll1$ and
$U\gg1$ to the extent possible, our primary interest is the regime in
between where $U$ is neither very small nor very large.  We have
picked functional forms for the various terms in the fitting function
such that they go smoothly between the two limits and agree as closely
as possible with the numerical results.  This is relatively easy to
achieve when only one of the three parameters, $r$, $r_s$, $\eta$, is
varied and the other two are set to zero (see the results shown in
Figs. \ref{figWr}, \ref{figWrs}, \ref{figWeta}).  However, we are also
interested in variations of two (e.g., the arc phenomenon, Appendix C)
or even three of these parameters simultaneously.  Finding a
reasonable fitting function to handle these regimes is more
challenging.  The particular function described in \S \ref{weakscatt}
and \S \ref{strongscatt} could doubtless be improved, but it appears
to fit the numerical results adequately over the entire parameter
range of interest.

% Read in ``Summary''
%\include{summary}
\section{Summary}\label{sec:summary}

We provide in \S4 of this paper a fitting function for a fairly
general correlation, $W(U,r,r_s,\eta)$, that describes the statistics
of flux scintillations of compact radio sources.  The function has
been optimized for two source shapes, gaussian and tophat; the latter
may be appropriate for gamma-ray burst afterglows (Sari 1998).  We
also allow for different source sizes and observation frequencies for
the two flux measurements being correlated, which may again be useful
for interpreting afterglow observations.

We have not specifically allowed for the finite integration time or
finite bandwidth over which each flux measurement is made.  Finite
integration time causes any given flux correlation to correspond not
to a single value of $r$ but to a range of values, i.e.,
$W(U,r,r_s,\eta)$ is smeared in $r$ by a convolution.  Similarly,
finite bandwidth leads to smearing in $\eta$.  These effects can be
modeled by convolving the fitting function with the appropriate
broadening functions in $r$ and $\eta$.  Alternatively, they could be
incorporated directly into the fitting function itself through
additional factors (we have not attempted this here).  In practice, with
current technological limits, finite bandwidth is unlikely to be important
in the regimes of weak or moderate scattering ($U\lesssim$~a few).

The results we have presented are for a single thin scattering screen.
This is probably a reasonable model since the scattering regions in
the interstellar medium tends to be clumpy.  Therefore, the scattering
is dominated by one or at most a few distinct screens.  The model is
also specific to a Kolmogorov spectrum of fluctuations in the
scattering medium.  This again is not unreasonable (e.g., Armstrong et
al. 1981), although Boldyrev \& Gwinn (2003) have recently challenged
the usual assumption, which we have adopted, that the statistics of
these fluctuations are gaussian.

The most serious simplifying assumption in this work is that we have
taken the scattering to be isotropic, so that the scintillation
correlation function depends on separation $r$ but not on the
particular direction of $r$ transverse to the line-of-sight.  A number
of observations (e.g., Wilkinson et al. 1994; Dennett-Thorpe
\& de Bruyn 2003; and references therein) have shown that the
scattering irregularities in the interstellar medium are anisotropic,
presumably because blobs are elongated parallel to the local magnetic
field.  It would be useful to generalize the work described here to
include anisotropy.

\acknowledgments We thank Edo Berger, Shri Kulkarni, and Stan Flatt\'e
for useful discussions, and our anonymous referee for a detailed
report containing many very helpful suggestions.  This work was
supported in part by NSF grant AST-0307433.

\appendix

% Read in Appendix on derivation of 4th-order correlation integral
%\include{appendixA}
\section{Derivation of the flux correlation integral}\label{AppA}

The goal of this appendix is to derive the expression \eqref{G4final}
for the fourth-order coherence function in the special case that the
points coincide in two pairs, so that it reduces to a flux
correlation.  
%Codona et al. (1986) 
\citet{codona86} have given a similar result for
frequency-independent refractive fluctuations, \emph{e.g.} atmospheric
rather than interstellar turbulence.  GN89 addressed the plasma case
but assumed small frequency differences, an approximation that is not
made here.

With $\bb=0$, eq.~(2.5.4) of GN89 becomes
\begin{eqnarray}\label{G4int}
\Gamma_4(\bzero;\br;\nu_1,\nu_2)&=&\frac{1}{(\lambda_1\lambda_2 z^2)^2}
\int d\bx_1\int d\bx_2\int d\bx_3\int d\bx_4\nonumber\\
&\phantom{=}&\exp\left\{i\pi\left[\frac{\bx_1^2-\bx_3^2}{\lambda_1 z}
-\frac{(\bx_2-\br)^2-(\bx_4-\br)^2}{\lambda_2 z}\right]\right\}\nonumber\\
&\phantom{=}&\exp\left\{-\half\left\langle\phi(\bx_1,\lambda_1)
-\phi(\bx_3,\lambda_1)-\phi(\bx_2,\lambda_2)+\phi(\bx_4,\lambda_2)
\right\rangle\right\}.
\end{eqnarray}
Here $\lambda=c/\nu$ and $z$ is the effective distance to the thin
screen, which impresses a phase shift $\phi(\bx,\lambda)$ on rays of
wavelength $\lambda$ that strike the screen at $\bx$.  The screen is
parallel to the observer's plane but separated from it by distance
$z_{\rm screen}$, while $z\equiv(z_{\rm screen}^{-1} +z_{\rm
source}^{-1})^{-1}$, where $z_{\rm source}$ is the distance from the
screen to the source.  Since $\phi$ involves the departure of the
refractive index from unity due to free electrons all along the path
of propagation, we assume that the $\phi\propto\lambda$.  Also,
$\phi(\bx,\lambda)$ is proportional to the column density of
electrons, which is stochastic.  The mean-square phase differences
impressed upon parallel paths meeting the screen at $\bx_1$ and
$\bx_2$ is
\begin{eqnarray}\label{strucdef}
D(\bx_1-\bx_2;\lambda)&\equiv&
\left\langle\left[\phi(\bx_1,\lambda)-\phi(\bx_2,\lambda)\right]^2\right\rangle
\nonumber \\
D(\bx_1-\bx_2;\lambda')&=& \left(\frac{\lambda'}{\lambda}\right)^2
D(\bx_1-\bx_2;\lambda)\,,
\end{eqnarray}
where $\langle\ldots\rangle$ is an appropriate statistical average.
Let $\bar\lambda\equiv(\lambda_1\lambda_2)^{1/2}$,
$\rho\equiv(\lambda_2/\lambda_1)^{1/2}$, $\phi_i\equiv\phi(\bx_i,\bar\lambda)$,
and $D_{ij}\equiv D(\bx_i-\bx_j,\bar\lambda)$.  Since 
$\langle\phi_i\phi_j\rangle=\sigma^2-\half D_{ij}$, where $\sigma^2$
is the variance of $\phi$,
the second exponential in eq.~\eqref{G4int} is
\begin{equation}\label{correxp}
\exp\left\{-\frac{1}{2}\left[\rho^{-2}D_{13}+\rho^2 D_{24} +D_{12}+D_{34}
- D_{14}-D_{23}\right]\right\}.
\end{equation}
Next, multiply \eqref{G4int} by
\[
\int d\bc~\delta\left(\bc-\frac{1}{4}\sum_{i=1}^4\bx_i\right)=1.
\]
Then shift variables $\bx\to\bx+\bc$, which has no effect on \eqref{correxp}
but multiplies the first exponential in \eqref{G4int} by
\[
\exp\left[2\pi i{\bf c\cdot}\left(\frac{\bx_1-\bx_3}{\lambda_1 z}-
\frac{\bx_2-\bx_4}{\lambda_2 z}\right)\right]\,.
\]
After integration over $\bc$, the net effect of these steps is to insert the
delta functions
\[
\delta\left(\frac{\bx_1+\bx_2+\bx_3+\bx_4}{4}\right)\,
\delta\left(\frac{\bx_1-\bx_3}{\lambda_1 z}-\frac{\bx_2-\bx_4}{\lambda_2 z}\right)
\]
into the integrand of \eqref{G4int}.  So, only two combinations
of the $\bx$s are independent.  It is convenient to choose these as
\begin{eqnarray*}
\bq&\equiv&\half(1+\rho^2)(\bx_1-\bx_3)~=~\half(1+\rho^{-2})(\bx_2-\bx_4)\\
\bs&\equiv& \half(\bx_1+\bx_3-\bx_2-\bx_4)\,,
\end{eqnarray*}
because the Fresnel (\emph{i.e.} first) exponential in \eqref{G4int} then
becomes
\[
\exp\left[\frac{4\pi i}{(\lambda_1+\lambda_2)z}{\bf q\cdot}
(\bs+\br)\right]
\]
After integrating out the two delta functions and expressing the remaining
integrals in terms of $\bq$ \& $\bs$, one has
\begin{eqnarray}\label{G4final}
\Gamma_4(\bzero;\br;\nu_1,\nu_2)&=& \frac{1}{(2\pi \rF^2)^2}\int\, d\bq
\exp\left(\frac{i\br\cdot\bq}{\rF^2}\right)
\int d\bs\, \exp\left(\frac{i\bq\cdot\bs}{\rF^2}\right)
\nonumber\\
&\times& 
\exp\left\{-\half\left[\frac{1-\eta}{1+\eta}D((1-\eta)\bq)
+\frac{1+\eta}{1-\eta}D((1+\eta)\bq)\right.\right.\nonumber\\
&&\qquad\left.\left.\vphantom{\frac{1-\eta}{1+\eta}}
+D(\bs-\eta\bq)+D(\bs+\eta\bq) -D(\bs-\bq)-D(\bs+\bq)\right]\right\}\,,
\end{eqnarray}
in which the structure functions (the $D$s)
are evaluated at the geometric mean $\sqrt{\lambda_1\lambda_2}$
of the two wavelengths, whereas the
Fresnel scale is defined with the arithmetic mean, eq.~\eqref{rFdef},
and $\eta$ is the dimensionless frequency difference \eqref{etadef}.
Equation \eqref{G4final} agrees with eqs.~(17)-(18) of Codona et al. (1986), apart from
what appear to be minor typographical errors\footnote{The term $1+(\delta k/2\bar k)$
should read $1-(\delta k/2\bar k)$ and {\it vice versa}.}
and for changes to allow for the $\lambda^2$ dependence of the plasma refractive index.

% Read in Appendix with asymptotic analytical formulae
%\include{appendixB}
\section{Asymptotic Results for the Flux Correlation}\label{AppB}

The bulk of this Appendix is devoted to analytical results for the
flux correlation in various asymptotic regimes.  Since the following
presentation is rather dense, we begin with an index to the main
results:
\begin{itemize}
\item \underline{Weak scintillation ($U\ll 1$)}:  Eqs.~\eqref{Wweak}, \eqref{varweak}.

\item \underline{Strong refractive scintillation}: 
Eqs.~\eqref{Wtref}-\eqref{Wref0}.

\item \underline{Strong diffractive scintillation}:
        \begin{itemize}
        \item {\it Total flux variance (\emph{i.e.}, $\br=0=\eta$):}
                                                 Eq.~\eqref{totvar}
        \item {\it Diffractive correlation at $r=0$ but $\eta\ne0$:}
                                         Eqs.~\eqref{Wfactored}-\eqref{fat0}.
        \item {\it Diffractive correlation at $r\ne0$ \& 
                $\eta\gg U^{-2/\alpha}$:} Eq.~\eqref{etanonzero}.
        \end{itemize}
\end{itemize}

\S \ref{subsec:finitesource} discusses the effect of a finite source size.

\subsection{Weak scattering ($U\ll 1$)}\label{subsec:weak}

The idea here is to take $\exp(-G/2)\approx 1-G/2$ in \eqref{Wtdef}.
This is justified because on the one hand,
$|G|\lesssim O\left(Uq^\alpha\right)$ since $G\to 0$ as $s/q\to\infty$;
and on the other hand, $Uq^\alpha\ll 1$ unless $q\gg1$,
in which case 
the rapidly oscillating factor $\exp(i\bq\cdot\bs)$ suppresses the
integral anyway:
\begin{eqnarray}\label{Wtexpand}
\Wt(\bq,\eta)&\approx& (2\pi)^2\delta(\bq)~-~
\half U\int d\bs\,e^{i\bq\cdot\bs}
\left[
\left|\bs-\eta\bq\right|^\alpha+\left|\bs+\eta\bq\right|^\alpha-
\left|\bs-\bq\right|^\alpha-\left|\bs+\bq\right|^\alpha\right]\nonumber\\
&& ~+~O(U^2).
\end{eqnarray}
The delta function simply reflects the fact that the mean flux is
nonzero.
In the remaining integral, each term has the form
\[
\int d\bs\,e^{i\bq\cdot\bs}\left|\bs-\ba\right|^\alpha=
e^{i\bq\cdot\ba}\int d\bt\,e^{i\bq\cdot\bt}
\left|\bt\right|^\alpha= 2\pi e^{i\bq\cdot\ba}\int\limits_0^\infty
t^{\alpha+1}J_0(qt)\,dt.
\]
The final integral is divergent, but for $-1<\alpha<-1/2$ it would be
\citep[\S 11.4.16]{AS}
\begin{equation}\label{bessel}
2\pi\int\limits_0^\infty t^{\alpha+1}J_0(qt)\,dt = - 2^{\alpha+2}
\Gamma^2(\alpha/2+1)\sin(\pi\alpha/2) q^{-\alpha-2}.
\end{equation}
We assume that this formula can in fact be used for all
$\alpha$.  Presumably the result could be justified
by inserting a slowly decreasing smooth function of $s$
into the original integrand \eqref{Wtdef}, for example $\exp(-\epsilon
s^2)$, and then taking the limit $\epsilon\to0$.  Such a convergence
factor is implicit in the Fresnel approximation to physical optics
since the phase screen is actually only a local approximation to a
closed surface enveloping the observer and therefore has finite area.
The final result is
\begin{eqnarray}\label{Wtweak}
\Wt(\bq;\eta,U)&\approx&
2^{\alpha+2}\Gamma^2(\alpha/2+1)\sin(\pi\alpha/2)\,U q^{-\alpha-2}
\left[\cos(\eta q^2)-\cos(q^2)\right] \nonumber \\
&& + O(U^2),\qquad q\ne0.
\end{eqnarray}
The flux correlation \eqref{Wdef} works out to
\begin{eqnarray}\label{Wweak}
W(\br;\eta,U)&\approx& 1+~2^\alpha\Gamma(\alpha/2+1)U\left[f(1,r)-f(\eta,r)
\right]~+O(U^2),\\[2ex]
\mbox{where}~~f(\eta,r)&\equiv& \eta^{\alpha/2}
\Re\left[e^{i\pi\alpha/4}M(-\alpha/2,1,ir^2/4\eta)\right]\,;\nonumber\\[1ex]
M(a,1,z) &\equiv& \sum\limits_{n=0}^\infty
\frac{\Gamma(a+n)}{\Gamma(a)(n!)^2}\, z^n ,\nonumber
\end{eqnarray}
is a confluent hypergeometric function \citep{AS}, and
$\Re$ denotes the real part.
The series for $M$ converges at all $z$.
The ``$1+$'' on the first line above represents the
square of the mean flux; we write $\Delta W\equiv W-1$.
The correlation at zero spatial separation is
particularly simple:
\begin{eqnarray}\label{varweak}
\Delta W(\bzero;\eta,U)
&\approx& 2^\alpha\Gamma(\alpha/2+1)\cos(\pi\alpha/4)\left(1-\eta^{\alpha/2}
\right)U~+~O(U^2)\\[1ex]
&\to& 0.7729\left(1-\eta^{\alpha/2}\right)U~+~O(U^2)
~~\mbox{at}~\alpha=5/3.\nonumber
\end{eqnarray}

\subsection{Strong scattering}\label{subsec:strong}

\subsubsection{Refractive regime}\label{subsubsec:ref}

Even when $U\gg 1$, the weak-scattering formula \eqref{Wtweak} is valid
if $q\ll\qref$,
\begin{equation}\label{qrefdef}
\qref\equiv U^{-1/\alpha}.
\end{equation}
Actually, one can extend the result to $q\sim\qref$ as follows.
The integral over $\bs$ is dominated by $s\sim 1/q$, where the
oscillating exponential cuts it off; in this range,
\begin{equation}\label{Gexpand}
G\approx (1-\eta^2) U q^{\alpha}\left[\alpha+\alpha(\alpha-2)\cos^2\theta
\right]\left(\frac{q}{s}\right)^{2-\alpha}~+~O\left(Uq^4 s^{\alpha-4}\right),
\qquad{s\gg q}\,,
\end{equation}
where $\cos\theta$ is the angle between $\bq$ and $\bs$.  Since we
assume $\alpha<2$, it follows that $G\ll Uq^\alpha\lesssim 1$, so that we may
again evaluate \eqref{Wtdef} by expanding $\exp(-G/2)$ to first order 
in its argument.  In contrast to the previous section, however,
we retain the prefactor $\exp(-F/2)$ because it cuts off the refractive
spectrum sharply at $q\sim\qref$:
\begin{equation}\label{Wtref}
\Wt_{\rm ref}(\bq)\approx 2^{\alpha+1}\sin(\pi\alpha/2)\Gamma^{\,2}(\alpha/2+1)
(1-\eta^2)Uq^{2-\alpha}\exp[-F(q;\eta,U)/2],\qquad q\lesssim\qref.
\end{equation}
Integrating over the direction of $\bq$ so that 
$\exp(\bq\cdot\bs)\to 2\pi J_0(qr)$, expressing $J_0(z)$ by its power series,
and integrating term by term, one finds
\begin{eqnarray}\label{Wref}
\Delta W_{\rm ref}(r;\eta,U)&\approx&
\frac{2^{\alpha-1}\Gamma(\alpha/2\,+1)\Gamma(4/\alpha\,-1)}
{\Gamma(-\alpha/2\,+1)}\,\frac{(1-\eta^2)U}{U_\eta}
\, U_\eta^{-2(2-\alpha)/\alpha}\nonumber\\[2ex]
&\times&\sum\limits_{n=0}^\infty\frac{\Gamma[(4+2n)/\alpha
\,-1]}{\Gamma(4/\alpha-1)\,(n!)^2}\,
\left(-\frac{r^2}{4U_\eta^{2/\alpha}}\right)^n.
\end{eqnarray}
We have introduced the abbreviation
\begin{equation}\label{Ueta}
U_\eta\equiv \frac{(1-\eta)^{\alpha+2}+(1+\eta)^{\alpha+2}}{2(1-\eta^2)}
\,U\,.
\end{equation}
Unfortunately the power series on the second line of \eqref{Wref}
is not any sort of hypergeometric function, 
but it converges for all $r$ if $\alpha>1$.
In particular,
\begin{equation}\label{Wref0}
\Delta W_{\rm ref}(0;\eta,U)\approx
0.2380 (1-\eta^2)\frac{U}{U_\eta}\,U_\eta^{-2/5},\qquad \alpha=5/3.
\end{equation}

\subsubsection{Diffractive regime}\label{subsubsec:diff}

If the dimensionless frequency difference $\eta\lesssim U^{-2/\alpha}$,
there is also a diffractive contribution to $W$ and $\Wt$.
This comes from $q\gg\qref\sim s$ in the integral \eqref{Wtdef}.
Consider first $\eta=0$. Then
\[
\half(F+G)\approx Us^\alpha\,\left[1~+~O(s/q)^{\alpha-2}\right].
\]
Keeping only the leading order term gives
\begin{equation}\label{Wdiffint}
\Delta W_{\rm diff}(\br;0,U)\approx\int\frac{d\bq}{(2\pi)^2}e^{i\br\cdot\bq}
\int d\bs\, e^{i\bq\cdot\bs}\exp\left(-U s^\alpha\right).
\end{equation}
But this is just a Fourier transform followed by its inverse.  So to
leading order,
\begin{equation}\label{Wdiff}
\Delta W_{\rm diff}(\br;0,U)\approx\exp\left(-U r^\alpha\right).
\end{equation}
We have written $\Delta W_{\rm diff}$ rather than $W$ because \eqref{Wdiffint}
applies only to $q\gg\qref$ and therefore omits both the refractive
contribution  \eqref{Wref} and the square of the mean flux,
both of which stem from from $q\lesssim\qref$.  The total flux correlation
is $1+\Delta W_{\rm ref}+\Delta W_{\rm diff}$.  

At the current level of approximation \eqref{Wdiff}, 
the contribution of diffractive scintillation to the flux variance
(the correlation at $\br=0$) is simply unity, \emph{i.e.}, equal
to the square of the mean flux, whereas the contribution from refraction
scales as $U^{-2(2-\alpha)/\alpha}$.  We can refine
the diffractive variance by the following trick.  With $\br=0$ and
$\eta=0$, the integral \eqref{Wdef} is symmetrical in $\bs$ \& $\bq$:
\[
W(\bzero;0,U)=\frac{1}{(2\pi)^2}\int d\bq\int d\bs\,
e^{i\bq\cdot\bs}\exp\left\{U\left[\half|\bs+\bq|^\alpha+\half|\bs-\bq|^\alpha
-s^\alpha-q^\alpha\right]\right\}\,.
\]
According to \S\ref{subsubsec:ref}, the contribution from $s\gg q$ is
$1+\Delta W_{\rm ref}(\bzero;0,U)$.  Symmetry
demands that the contribution from $s\ll q$ must be the same:
\begin{equation}\label{diffvar}
\Delta W_{\rm diff}(\bzero;0,U)\approx~1+\Delta W_{\rm ref}(\bzero;0,U)\,,
\end{equation}
and therefore the total flux variance in a very narrow frequency band is
\begin{equation}\label{totvar}
\frac{\langle F^2\rangle}{\langle F\rangle^2}~-1~=
~1+2\Delta W_{\rm ref}(\bzero;0,U)\,.
\end{equation}
The factor of two reflects equal contributions from the refractive ($s\gg q$)
and diffractive ($s\ll q$) regimes; but
this symmetry should not obscure the fact that
the diffractive and refractive contributions decorrelate on widely
different lengthscales $r_{\rm diff}\sim U^{-1/\alpha}$ 
[eq.~\eqref{Wdiff}] and
$r_{\rm ref}\sim U^{+1/\alpha}$ [eq.~\eqref{Wref}], respectively.

Let us now consider nonzero frequency differences, $\eta\ne0$.
The most sensitive terms are the
structure functions $U|\bs\pm\eta\bq|^\alpha$; other appearances of
$\eta$ are smaller by at least $O(\eta^{2-\alpha})$:
\begin{equation}\label{Wwitheta}
\Delta W_{\rm diff}(\br;\eta,U)\approx\int\frac{d\bq}{(2\pi)^2}
e^{i\br\cdot\bq}\int d\bs\,e^{i\bq\cdot\bs}\exp\left[-\half U\left(
|\bs+\eta\bq|^\alpha+|\bs-\eta\bq|^\alpha\right)\right].
\end{equation}
This reduces to \eqref{Wdiffint} at $\eta=0$.
With the change of variables
\begin{eqnarray}\label{changevars}
\bs&=&\left(\frac{\eta}{2}\right)^{1/2}(\bu+\bv),\qquad
\bq=\left(\frac{1}{2\eta}\right)^{1/2}(\bu-\bv),\nonumber\\
\br&=&(2\eta)^{1/2}\brho,\qquad\mbox{and}\quad
V = \half(2\eta)^{\alpha/2}U,
\end{eqnarray}
the double vector integral can be factored into the product of independent
integrations over $\bu$ and $\bv$ that are complex conjugates:
\begin{eqnarray}\label{Wfactored}
\Delta W_{\rm diff}(\br;\eta,U) &\approx& 
\left|f(\rho,V)\right|^2,\\[1ex]
\mbox{where}~~f(\rho,V) &\equiv& 
\int\frac{d\bu}{2\pi}e^{-i\brho\cdot\bu-iu^2/2-Vu^\alpha}
~=~\int\limits_0^\infty J_0(\rho u)\exp\left(-iu^2/2-Vu^\alpha\right)\,udu.
\nonumber
\end{eqnarray}
[The object $f(\rho,V)$ is equivalent to the two-frequency, two-position
second moment of the electric field given by eq.~(50) of \citet{lam99}, who
also cite previous work on the latter quantity.  It is well known that
fourth moments of the field, such as $W$, can usually be approximated by
squares of second moments in very strong scattering 
\citep[cf. \emph{e.g.}][]{codona86}.]
The monochromatic limit $\eta\to0$ is recovered by taking
$V\to0$ while keeping $(2V)^{1/\alpha}\rho=U^{1/\alpha}r$ constant.
We have no general closed-form expression for $f(\rho,V)$, but at $r=0$,
\begin{eqnarray}\label{fat0}
f(0,V)&=&
-i\sum\limits_{n=0}^\infty\frac{\Gamma(n\alpha/2+1)}{n!}
\left(-e^{-i\pi\alpha/4}2^{\alpha-1}\eta^{\alpha/2}U\right)^n\,\nonumber\\
&\sim& \frac{2^{2/\alpha}}{2\alpha\eta U^{2/\alpha}}
\sum\limits_{n\ge0}\frac{\Gamma[2(n+1)/\alpha]}{n!}
\left(-\frac{i}{2}V^{-2/\alpha}\right)^n\,.
\end{eqnarray}
% Added by JJG 8/19/05
Looking at the first two terms of this sum, one sees that for small $\eta$,
the intensity decorrelates as
\[
W(\bzero;0,0)-W(\bzero;\eta,0)\approx 2^\alpha\Gamma\left(\frac{\alpha+2}{2}\right)
\cos\left(\frac{\pi\alpha}{4}\right)\eta^{\alpha/2} U
\]
in agreement with eq.~(21) of Shishov et al. (2003) if one assumes that the scattering,
which they take to be distributed along the line of sight, is instead
concentrated on a thin screen at distance $R/2$, and if the
term $\cos\frac{n-2}{4\pi}$ should have read $\cos\frac{(n-2)\pi}{4}$ therein.
% end of addition 8/19/05
The second sum in \eqref{fat0} is asymptotic and shows that
\begin{equation}\label{largeV}
\Delta W_{\rm diff}(\bzero;\eta,U)\approx 2^{-4(\alpha-1)/\alpha}
\Gamma^{\,2}(2/\alpha\,+1)\left(\eta U^{2/\alpha}\right)^{-2}\quad
\mbox{if}~\eta\gg U^{-2/\alpha}.
\end{equation}
From either sum, one sees that the decorrelation bandwidth is
$\eta\sim U^{-2/\alpha}=\qref^2$.

Let us now consider the dependence on spatial lag $\br$.
For $\eta\lesssim\qref^2$, this is
approximately the same as for $\eta=0$, eq.~\eqref{Wdiff}.  
For $\eta\gg\qref^2$, the term $iu^2/2\ll Vu^\alpha$ in \eqref{Wfactored}
except where the integrand is negligibly small anyway.  By omitting
$iu^2/2$ from the exponential and rescaling the dummy
variable $u\to V^{-1/\alpha} t$, one can obtain
\begin{eqnarray}\label{etanonzero}
\Delta W_{\rm diff}(\br;\eta,U)&\approx& \Delta W_{\rm diff}(\bzero;\eta,U)
\left|g\left[\frac{r}{2\eta (U/2)^{1/\alpha}}\right]\right|^2,
\\[2ex]
\mbox{where}~~g(z) &=&
\frac{\alpha}{\Gamma(2/\alpha)}\int\limits_0^\infty J_0(zt)e^{-t^\alpha}
\,tdt.\nonumber\\
&=&\frac{1}{\Gamma(2/\alpha)}\sum\limits_{n=0}^\infty
\frac{\Gamma[2(n+1)/\alpha]}{(n!)^2}
\left(-\frac{z^2}{4}\right)^n\,.\nonumber
\end{eqnarray}
[Mathematically $g(z)$, is essentially the same as the function 
$Q_{\rm PD}(\tau)$ given in eq.~(54) of \citet{lam99}, but the latter
is interpreted physically as a pulse-broadening profile at zero spatial
lag.]
To be consistent with the assumption $\eta\gg U^{-2/\alpha}$, one
ought to use \eqref{largeV} for $\Delta W_{\rm diff}(\bzero;\eta,U)$
in the above formula, but one might want to use the more accurate
value $|f(0,V)|^2$ as given by \eqref{fat0} instead.

\subsection{Finite Source Size}\label{subsec:finitesource}

Equations (\ref{finitesource})--(\ref{tophat}) show how a finite
source size modifies the flux correlation.  For zero separation
($r=0$), for instance, we have
\begin{equation}\label{Wfinitesource}
W(r=0,r_s\ne0) \approx {1\over2\pi}\int_0^{1/r_s} \widetilde W(q) qdq,
\end{equation}
where $r_s$ refers to the effective source defined in equation
(\ref{rsdef}), and we have made use of the fact that the Hankel
transform $\widetilde S(q)$ of a source with a finite size cuts off at
$q \sim 1/r_s$.  Using this result, we may estimate the effect of a
finite source size in the different scattering regimes.  In the
following, we assume for simplicity that $r=\eta=0$.

In weak scattering, there is a reduction in the flux variations when
the source size is greater than the Fresnel scale, i.e., $r_s > 1$ in
our units.  Thus, the relevant range of $\widetilde W(q)$ for
evaluating the integral in equation (\ref{Wfinitesource}) is $q <
1/r_s \lesssim 1$.  In this regime, $\widetilde W(q) \sim
q^{2-\alpha}$ (see equation \ref{Wtweak}), and hence we obtain
\begin{equation}\label{Wweakfinitesource}
W_{\rm weak}(r=0,r_s,\eta=0) \propto r_s^{\alpha-4}, \qquad r_s > 1.
\end{equation}

In strong refractive scintillation, finite source effects are felt
when the source size exceeds the refractive scale $r_{\rm
ref}\sim1/q_{\rm ref}$.  In this regime again $\widetilde W(q)$ varies
as $q^{2-\alpha}$ (see eq. \ref{Wtref}), and so we obtain
\begin{equation}\label{Wreffinitesource}
W_{\rm ref}(r=0, r_s,\eta=0) \propto (r_s/r_{\rm ref})^{\alpha-4},
\qquad r_s > r_{\rm ref}.
\end{equation}

Finally consider strong diffractive scintillation.  In the asymptotic
limit of very strong scattering ($U\gg1$), $\widetilde W(q)$ is
constant ($\widetilde W(q) \sim q^0$, white noise) upto $q \sim q_{\rm
diff}$ (e.g., Goodman \& Narayan 1985).  For a source size $r_s >
r_{\rm diff} \sim 1/q_{\rm diff}$, this gives
\begin{equation}
W_{\rm diff}(r=0, r_s,\eta=0) \propto (r_s/r_{\rm diff})^{-2},
\qquad r_s > r_{\rm diff}, ~U \gg 1.
\end{equation}
However, unless $U$ is extremely large, the next order term in
$\widetilde W(q)$ cannot be neglected, and it gives a correction that
causes $\widetilde W(q)$ to decrease with increasing $q$ (e.g., see
Fig. 1 of Goodman \& Narayan 1985, where the decrease is apparent even
for $U=10^4$).  As a result, the scaling due to a finite source size
is modified slightly to
\begin{equation}\label{Wdifffinitesource}
W_{\rm diff}(r=0, r_s,\eta=0) \propto (r_s/r_{\rm diff})^{-\beta},
\qquad r_s > r_{\rm diff}, ~\beta<2,
\end{equation}
where the exact value of $\beta$ depends on $U$; $\beta$ is distinctly
less than 2 when $U$ is not very large, but it tends to 2 as
$U\to\infty$.

\section{Parabolic arcs in secondary spectra}\label{AppC}

The dynamical spectrum of a scintillating source is the flux correlation
in a two-dimensional plane of time and frequency lag. Its Fourier transform
with respect to both arguments is called the \emph{secondary} \emph{spectrum}
$S_2(f_\nu,f_t)$.
When the interesting range of frequencies is small compared to the mean
frequency, $\Delta\nu\ll\nu$, and
when the correspondence between spatial and temporal lags is governed by
the transverse velocity $\bv_\perp$ of line of sight, $\br=\bv_\perp t$,
the secondary spectrum is related to the cross spectrum \eqref{Wtdef} by
\begin{equation}\label{S2def}
S_2(f_\nu,f_t)= \frac{\rF}{v_\perp}\int\frac{dq'}{2\pi}
\int \nu d\eta~\Wt\left(q_t\bvh+q'\beh,\eta\right)
e^{2\pi i f_\nu \nu\eta}\,,
\end{equation}
in which $v_\perp\equiv|\bv_\perp|$, and $\bvh$ \& $\beh$ are unit vectors
on the sky parallel \& perpendicular to $\bv_\perp$, respectively,
while $q_t\equiv 2\pi \rF f_t/v_\perp$ is the component of the wavenumber
parallel to the motion through the scintillation pattern.
The Fresnel scale \eqref{rFdef} appears here because $\Wt$ and $\bq$ have
been expressed in units where $\rF=1$: that is, the physical wavenumber is $\rF\bq$.
The integration over $q'$ reduces the 2D cross-spectrum $\Wt$ to a 1D cross-spectrum.

Secondary spectra often show a concentration of power along parabolic
ridges $f_\nu=\pm a f_t^2$ \citep{stinebring01}.
\citet{cor04} have given a theoretical explanation of this phenomenon, according
to which the coefficient $a=z\lambda^2/(cv_\perp^2)$, where $z$ is the effective
distance of the screen \eqref{zdef}.

The existence of parabolic arcs in strong diffractive scattering can be demonstrated
from equation \eqref{Wwitheta}, which is equivalent to
\begin{equation}\label{Wtwitheta}
\Wt_{\rm diff}(\bq;\eta,U)\approx\int d\bs\,e^{i\bq\cdot\bs}\exp\left[-\half U\left(
|\bs+\eta\bq|^\alpha+|\bs-\eta\bq|^\alpha\right)\right].
\end{equation}
This is a good approximation to the exact relations \eqref{Wtdef}-\eqref{Gdef}
when $\eta^2 U q^\alpha\ll 1$ (even though $Uq^\alpha$ may be large)
so that the neglected terms in the exponential approximately cancel.
For large $q$, the integral over $\bs$ is dominated by
the neighborhood of its branch points.  The leading-order contribution from
$\bs\approx\eta\bq$ can be obtained by setting
$|\bs+\eta\bq|^\alpha\to|2\eta\bq|^\alpha$ (because this part of the integrand
remains smooth in that neighborhood) and then expanding the remaining exponential
to first order in its argument; treating $\bs\approx-\eta\bq$ similarly, one has
\begin{eqnarray*}
\Wt_{\rm diff}(\bq;\eta,U)&\approx&
e^{-(U/2)|2\eta q|^\alpha}\left[
\int d\bs\,e^{i\bq\cdot\bs}\left(1-\frac{U}{2}
|\bs-\eta\bq|^\alpha\right)
~+~
\int d\bs\,e^{i\bq\cdot\bs}\left(1-\frac{U}{2}
|\bs+\eta\bq|^\alpha\right)\right]\\
&\approx&
2(2\pi)^2\delta(\bq)~+~K_\alpha Uq^{-\alpha-2}\,
\cos(\eta q^2)e^{-(U/2)|2\eta q|^\alpha}\,,
\end{eqnarray*}
where $-K_\alpha$ is the coefficient of $q^{-\alpha-2}$ in eq.~\eqref{bessel}.
The delta function must be discarded since $q$ is large.
When the remainder is used  in \eqref{S2def} to calculate the secondary spectrum,
the exponential in $U|2\eta q|^\alpha$ can be replaced by unity if
\begin{equation}\label{fnulim}
f_\nu\gg\frac{\rF}{\nu v_\perp}U^{1/\alpha}f_t\,,
\end{equation}
because the complex exponential will then effectively
limit $\eta$ to small values:
\begin{equation}\label{S2p}
S_2(f_\nu,f_t)\approx \half K_\alpha U\frac{r_F\nu}{v_\perp}\int\frac{dq'}{2\pi}
\left(q_t^2+q^{\prime\,2}\right)^{-(\alpha+2)/2}
\left[\delta\left(\nu f_\nu+\frac{q_t^2+q^{\prime\,2}}{2\pi}\right)~+~
\delta\left(\nu f_\nu-\frac{q_t^2+q^{\prime\,2}}{2\pi}\right)\right].
\end{equation}
The parabolic arcs can now be recognized in the arguments of these delta functions,
which are softened but not suppressed by the final integration over the transverse
wavenumber:
\begin{eqnarray}\label{S2final}
S_2(f_\nu,f_t)&\approx&(2\pi)^{-3/2}\left(\frac{2}{\pi}\right)^{\alpha/4}
\Gamma^2\left(\frac{\alpha+2}{2}\right)\sin\left(\frac{\pi\alpha}{2}\right)U
\frac{\sqrt{cz}}{v_\perp}\nonumber\\
&\times&~ \left(\frac{v_\perp^2}{\lambda z f_t^2}\right)^{(\alpha+2)/4}
\left(|f_\nu|-af_t^2\right)^{-1/2} H(|f_\nu|-af_t^2).
\end{eqnarray}
Here $a=\lambda^2 z/c v_\perp^2$ as before, and the Heaviside function
keeps the argument of the square root positive.  The Heaviside
function cuts off the power abruptly for $|f_t| > (|f_\nu|/a)^{1/2}$, and
the inverse square root concentrates power just inside this cutoff.
Both observations and numerical simulations confirm this general
behavior \citep{cor04}, and this is the origin of the arcs.  Plausibly
the arcs would be more prominent if the electron-density spectrum were
anisotropic with stronger scattering along $\bvh$ than along $\beh$,
because the integration \eqref{S2p} would have less of a smearing
effect.  Presumably also, arcs should be more prominent when
scattering is dominated by a thin ``screen'' rather than distributed
along the line of sight, since integration over $z$ would further
soften the final square root in eq.~\eqref{S2final}.  These points are
made by \citet{cor04}.

In strong scattering ($U\gg1$), the conditions for validity of the
final result \eqref{S2final} include not only eq.~\eqref{fnulim} but
also $q_t\equiv 2\pi f_t/v_\perp\gg U^{-1/\alpha}=q_{\rm diff}$ so
that the integration over $\bs$ is dominated by $s\ll q$.  However,
the same result obtains in weak scattering as well provided $2\pi f_t
\gg v_\perp/\rF$, since the same $\cos(\eta q^2)$ term appears in
eq.~\eqref{Wtref} and gives rise to delta functions in the $\eta\to
f_\nu$ transform of the 2D cross spectrum.  In fact, regardless of the
magnitude of the scintillation parameter $U$, it appears that the arcs
are due to weak large-angle scattering by inhomogeneities smaller than
those that are responsible for the main part of the scatter-broadened
image \citep{cor04}.



\end{document}